\documentclass[fleqn,usenatbib]{mnras}

\usepackage{newtxtext,newtxmath}
\usepackage[T1]{fontenc}

\DeclareRobustCommand{\VAN}[3]{#2}
\let\VANthebibliography\thebibliography
\def\thebibliography{\DeclareRobustCommand{\VAN}[3]{##3}\VANthebibliography}

\usepackage{xcolor}
\usepackage{textgreek}
\usepackage[utf8]{inputenc}
\usepackage[english]{babel}

\usepackage{amsmath}
\usepackage{bm}

\usepackage{hyperref}
\hypersetup{
    unicode, 
    colorlinks=true,
    linkcolor=linkcolor,
    citecolor=linkcolor,
    filecolor=linkcolor,
    urlcolor=linkcolor,
}
\usepackage{color,colortbl}
\definecolor{linkcolor}{rgb}{0.0,0.3,0.5}
\usepackage{tensind}
\tensordelimiter{?}
\usepackage{verbatim}
\usepackage[normalem]{ulem}
\usepackage{orcidlink}
\usepackage{soul}
\usepackage{amsmath}

\newcommand{\Msun}{M_\odot}
\newcommand{\md}{\mathrm{d}}

\newcommand{\bJ}{\bm{J}}
\newcommand{\Rg}{R_{\mathrm{g}}}
\newcommand{\nn}{\nonumber}

\newcommand{\mr}{\mathrm}

\renewcommand{\l}{\left}
\renewcommand{\r}{\right}
\graphicspath{ {./Figs/} } 
\title[Stochastic kicks for ISM-driven diffusion]{A stochastic kick recipe for interstellar medium-driven orbital diffusion}

\author[Winter-Granic et al.]{
Mila Winter-Granic,$^{1}$\thanks{E-mail: milawinter@princeton.edu}
Shaunak Modak,$^{2,3}$
Chris Hamilton,$^{1}$
Adrian~M.~Price-Whelan,$^{2}$
\\
$^{1}$Department of Astrophysical Sciences, Princeton University, 4 Ivy Lane, Princeton, NJ 08544, USA\\
$^{2}$Center for Computational Astrophysics, Flatiron Institute, 162 5th Avenue, New York, NY 10010\\
$^{3}$Canadian Institute for Theoretical Astrophysics, University of Toronto, 60 St. George Street, Toronto, ON M5S 3H8, Canada}

\begin{document}
\label{firstpage}
\pagerange{\pageref{firstpage}--\pageref{lastpage}}
\maketitle


\begin{abstract}
Stellar orbits in the Galactic disc are subject to stochastic gravitational perturbations from interstellar medium (ISM) substructure. However, $N$-body and test-particle models of the Galaxy typically treat the ISM very crudely, or ignore it altogether. Meanwhile, state-of-the-art ISM simulations have recently shown that the classic Spitzer--Schwarzschild paradigm, in which the ISM's influence is modelled as a series of uncorrelated, impulsive velocity kicks from giant molecular clouds, must be revised. In this paper we show how to implement realistic ISM forcing when integrating disc star orbits. Our recipe takes as input just three action-space diffusion coefficients, and produces the correct ISM-driven diffusion properties across the full six-dimensional phase space, at a fraction of the cost of a full (magneto-)hydrodynamic simulation. We implement the method in the \texttt{gala} code and illustrate the ISM's stellar-dynamical effect in simple models of both the \textit{Gaia} Snail and the Hercules `tree ring' structure. We show that the finite correlation time of ISM fluctuations is crucial: realistic diffusion is actually \textit{less} destructive of fine-grained phase space features than if those fluctuations were temporally uncorrelated, as is commonly assumed. Beyond the ISM, our recipe applies to arbitrary diffusive forcing defined in action space, providing a way to directly and efficiently test how stochastic fluctuations reshape the phase-space structure of galactic discs.
\end{abstract}


%%%%%%%%%%%%%%%%%%%%%%%%%%%%%%%%%%%%%%%%%%%%%%%%%%

%%%%%%%%%%%%%%%%% BODY OF PAPER %%%%%%%%%%%%%%%%%%

%%%%%%%%%%%%%%%%%%%%%%%%%%%%%%%%%%%%%%%%%%%%%%%%%%
%%%%%%%%%%%%%%%%%%%%%%%%%%%%%%%%%%%%%%%%%%%%%%%%%%
\section{Introduction}
\label{sec:intro}
%%%%%%%%%%%%%%%%%%%%%%%%%%%%%%%%%%%%%%%%%%%%%%%%%%
%%%%%%%%%%%%%%%%%%%%%%%%%%%%%%%%%%%%%%%%%%%%%%%%%%

The orbits of stars in our Galactic disc are continually perturbed by stochastic fluctuations in its gravitational potential. These fluctuations might arise from non-axisymmetric structures such as transient spiral arms, from dark matter subhaloes, from density fluctuations in the interstellar medium (ISM), and so on. Over timescales much longer than an orbital period (i.e., hundreds of Myr or longer) these fluctuations drive the diffusion of stars through phase space \citep{binney_lacey1988,sellwood2014}. The resulting `orbital transport' can be helpfully divided into three pieces: radial heating (changes in a star's radial action $J_R$), vertical heating (changes in its vertical action $J_z$), and radial migration (changes in its angular momentum $J_\varphi$, or equivalently its 
guiding radius $\Rg$).

Understanding orbital heating and migration is crucial for interpreting Galactic disc observations. There are two main reasons for this. First, heating and migration underlie bulk disc properties like the age-velocity dispersion relation, the radial metallicity profile and the scatter around it, etc., that evolve over billions of years \citep[e.g.,][]{sellwood_binney2002, mackereth2019, frankel2020, Zhang2026, Modak2026residuals}. These are properties tied to the disc's axisymmetric, mean-field distribution function (DF), $f_0$. Secondly, in the era of \textit{Gaia} we are also able to measure fine-grained---and perhaps shorter-lived---features in the Galactic disc's phase space such as the radial velocity wave \citep{friske_schonrich2019}, the vertical phase space snail \citep{antoja2018}, and the tree ring metallicity structure in the Hercules stream \citep{chiba_tree-ring_2021}. These features are tied to the non-axisymmetric, angle-dependent \textit{fluctuations} in the disc DF, $\delta f \equiv f - f_0$. Orbital transport has the tendency to blur or erase fine-grained structure in $\delta f$, so must be accounted for when interpreting these features. Despite this, very many efforts at modelling the Milky Way's history and comparing it to \textit{Gaia} data have tended to include these stochastic effects in a crude way or simply ignored them altogether \citep[see, e.g.,][for a recent review]{hunt_vasiliev2025}.

In this paper we will focus on the orbital heating and migration driven by ISM density fluctuations (although as we will discuss later, our recipe applies to disc stars undergoing any diffusion process defined in action space). Historically, the effect of the ISM on stellar orbits has usually been modelled following the classic paradigm of  \citet{spitzerschwarzchild_51,spitzerschwarzchild_53}. These authors proposed that disc stars receive a series of impulsive kicks from giant molecular clouds (GMCs), usually modelled as long-lived Plummer spheres on circular orbits or similar. The GMC picture leads to a series of predictions for the orbital heating and migration of disc stars, and hence for observables like the age-velocity dispersion, the scale height of various disc populations, and so on \citep[e.g.,][]{lacey1983kinematical,lacey1984,hanninen_flynn2002,aumer2016}. This dominant paradigm has also been used to understand the blurring/erasure of phase-mixing structures in the \textit{Gaia} data \citep[e.g.,][]{chiba2025galactic,chiba_snail2025,tremaine_snail2023}.

However, we now know that the ISM does not consist of long-lived spheroidal clouds, but rather is a turbulent medium exhibiting a hierarchy of extended structures that span a broad range of spatial and temporal scales. \citet{modak_interstellar_2026} recently revisited the ISM-driven transport problem by integrating stellar orbits in the fluctuating gravitational potential of the TIGRESS-NCR magnetohydrodynamic simulations of the local ISM \citep{Kim2023}. They found that the predictions of the classic GMC-based picture did not hold. Instead, for a given amount of radial migration, a realistic ISM heats stellar orbits significantly less than the GMC picture would suggest. Briefly, the key differences between the GMC-based picture and the realistic ISM are that (i) transport is driven predominantly by fluctuations on spatial scales of $\sim 600\,$pc, and (ii) fluctuations on these scales are correlated over finite timescales of $\sim 70\,$Myr \citep{Modak2026characterizing}. \citet{modak_interstellar_2026} showed that an action-based kinetic theory that accounted for these key differences was able to reproduce the simulation results \citep[see also][]{hamilton_fouvry2024,hamilton_galactokinetics_2026}. Moreover, the results are not particular to the TIGRESS-NCR setup but are instead generic; any Gaussian random field of gravitational potential fluctuations dominated by the same characteristic time and length scales drives the same type of diffusion (Hamilton, Modak \& Winter-Granic, in preparation).

In summary, we know that ISM-driven diffusion is important to model realistically, and we know how to describe it as an action-space diffusion process. But implementing this diffusion in standard numerical simulations is a separate challenge. While actions are ideal for isolating the changes to orbits caused by perturbations (because in an unperturbed axisymmetric potential they are conserved), orbit integrations \citep[e.g.,][]{bovy2015,pricewhelan2017,vasiliev2019} and $N$-body simulations \citep[e.g.,][]{roskar2008,aumer2016,teppergarcia2024} evolve stars in position-velocity space. Therefore, to implement ISM-driven transport in simulations (without running self-consistent multi-physics MHD calculations at the same time), we need a recipe that translates the action-based kinetic theory of \citet{modak_interstellar_2026} and Hamilton, Modak \& Winter-Granic (in preparation) into position-velocity space.

In this paper we develop such a recipe for stars in the Galactic disc. We first derive the drift and diffusion coefficients in position and velocity that reproduce a given action-space diffusion process. We then describe a procedure to model this diffusion via stochastic (Langevin) kicks that can be coupled to standard orbit integrators, and implement the method in \texttt{gala} \citep[v1.13;][]{pricewhelan2017}. Our recipe takes as its input just three action-space diffusion coefficients $D_{J_\varphi J_\varphi}, D_{J_R J_R}$ and $D_{J_z J_z}$, yet is able to reproduce the full six-dimensional phase space ensemble-averaged diffusion behaviour measured from direct orbit integrations through the TIGRESS-NCR ISM simulations, at a fraction of the computational cost.

The rest of this paper is organized as follows. In \S\ref{sec:framework} we write down the basic angle-action to position-velocity mapping we use for disc stars on epicyclic orbits, and derive the general stochastic-kick recipe that works for arbitrary diffusion processes defined in action space. In \S\ref{sec:ism} we specialize to the ISM, recovering the local simulation results of \cite{modak_interstellar_2026} and extending them to a global setting. In \S\ref{sec:discussion} we explain why kicks in velocity alone do not suffice for our recipe, illustrate how ISM-driven diffusion affects two structures in the Milky Way's disc (the \textit{Gaia} phase spiral and the metallicity tree rings in the Hercules stream), and discuss the implications and possible extensions of our work. We summarize in \S\ref{sec:summary}.

%%%%%%%%%%%%%%%%%%%%%%%%%%%%%%%%%%%%%%%%%%%%%%%%%%
%%%%%%%%%%%%%%%%%%%%%%%%%%%%%%%%%%%%%%%%%%%%%%%%%%
\section{General framework}
\label{sec:framework}
%%%%%%%%%%%%%%%%%%%%%%%%%%%%%%%%%%%%%%%%%%%%%%%%%%
%%%%%%%%%%%%%%%%%%%%%%%%%%%%%%%%%%%%%%%%%%%%%%%%%%

%%How should one kick a star in a real-space numerical simulation 
% so that its actions diffuse at a prescribed
%rate?
%To answer this question we first relate the actions of a star to its position and velocity
%(\S\ref{sec:orbits}), and then describe diffusion in action space (\S\ref{sec:action_diffusion}).
%Then we write down the kick prescription (\S\ref{sec:kicks}), though we relegate
% the explicit formulae to 
% Appendix \ref{app:prescriptions}.

\subsection{Epicyclic orbits}
\label{sec:orbits}

We describe the position and velocity of a star using standard cylindrical coordinates $\bm{r}=(\varphi, R, z)$ and $\bm{v}=(v_\varphi,v_R, v_z)$ centred on the Galactic centre.
We transform to angle-action variables $(\bm{\theta},\bm{J})$ using the epicyclic approximation, in which stellar motions may be decomposed into circular motion at the guiding radius $\Rg$, small (with respect to $\Rg$) radial oscillations about $\Rg$, and small (with respect to the potential scale height) vertical oscillations about the midplane $z=0$ with frequencies
\begin{align}
    \kappa^2 &= \frac{\partial^2\phi_\mr{eff}}{\partial R^2}\bigg|_{R=\Rg, z = 0} = \l(R\frac{\md\Omega^2}{\md R}+4\Omega^2\r)_{R=\Rg, z = 0}, \label{eq:kappa}\\
    \nu^2 &= \frac{\partial^2\phi_\mr{eff}}{\partial z^2}\bigg|_{R=\Rg,\,z=0}, \label{eq:nu}
\end{align}
respectively. Here, $\phi_\mr{eff}=\phi(R, z)+J_\varphi^2/(2R^2)$ is the effective potential, $\phi$ is the 
Galaxy's unperturbed axisymmetric potential, $J_\varphi$ is the angular momentum, and $\Omega(R)$ is the angular
frequency of a circular orbit of radius $R$. The azimuthal, radial and vertical actions $\bJ = (J_\varphi, J_R, J_z)$ are then
\begin{align}
        J_\varphi & \equiv R v_\varphi \equiv \Rg^2\,\Omega(\Rg), \label{eq:Jphi}\\
    J_R & \equiv \frac{v_R^2+\kappa^2(R-\Rg)^2}{2\kappa}, \label{eq:JR}\\
    J_z & \equiv \frac{v_z^2+\nu^2z^2}{2\nu}, \label{eq:Jz}
\end{align}
where the frequencies $\kappa$ and $\nu$ are evaluated at $\Rg$, and the second equality in Equation~\eqref{eq:Jphi} defines $\Rg$. The corresponding angles $\btheta = (\theta_\varphi, \theta_R, \theta_z)$ may then be calculated from the epicyclic mapping 
\begin{subequations}
    \label{eq:EOM}
    \begin{align}
        \varphi &= \theta_\varphi + \gamma\,\frac{a_R}{\Rg}\sin\theta_R, \\
        R &= \Rg - a_R\cos\theta_R, \\
        z &= -a_z\cos\theta_z, \\
        v_\varphi &= v_c(\Rg) + \Omega\,a_R\cos\theta_R, \\
        v_R &= \kappa\,a_R\sin\theta_R, \\
        v_z &= \nu\,a_z\sin\theta_z,
    \end{align}
\end{subequations}
where
\begin{equation}
    a_R \equiv \sqrt{\frac{2J_R}{\kappa}}, \quad a_z \equiv \sqrt{\frac{2J_z}{\nu}},
\end{equation}
are the amplitudes of the radial and vertical oscillations, $\gamma\equiv2\Omega/\kappa$, and $v_c(R)=R\,\Omega(R)$ is the circular velocity at radius $R$. This mapping, and everything we derive from it subsequently, are accurate to first order in both radial and vertical epicyclic amplitudes.

\subsection{Diffusion equation}
\label{sec:action_diffusion}

In the Galaxy's unperturbed axisymmetric potential the actions $\bJ$ are conserved, while the angles increase at the constant rates $\Omega$, $\kappa$ and $\nu$ respectively. Changes in the actions therefore isolate the effect of gravitational potential fluctuations, without the oscillatory changes in position and velocity that arise from the unperturbed motion. Thus, angle-action space is the natural place to measure the diffusion of orbits (and indeed to formulate theories of this diffusion). In this section we will focus only on modelling the diffusion of orbital actions, but we will see in \S\ref{sec:box} that for ISM-driven diffusion, our models also yield approximately correct diffusion in angles.

We can describe action-space diffusion with the following diffusion equation for the angle-averaged DF $f_0(\bJ,t)$:
\begin{equation}
    \label{eq:QL_diffusion}
    \frac{\partial f_0}{\partial t} = \frac{1}{2}\sum_{i,j}\frac{\partial}{\partial J_i}
    \l(D_{J_iJ_j}\,\frac{\partial f_0}{\partial J_j}\r),
\end{equation}
where the indices $i$ and $j$ run over $\varphi$, $R$ and $z$  (see, e.g., \citealt{binney_lacey1988, binney_tremaine2008}). The diffusion tensor
\begin{equation}
    D_{J_iJ_j}(\bJ) = \frac{\overline{\langle\Delta J_i\,\Delta J_j\rangle}}{\Delta t},
    \label{eqn:D_JJ}
\end{equation}
is the product of the changes in $J_i$ and $J_j$ per unit time driven by the fluctuations, averaged in two ways: angle brackets denote an average over many realizations of the fluctuations for a star with given actions and angles, and the overbar denotes a subsequent average over the orbital angles $\btheta$ at fixed actions $\bJ$. The action increments are measured over a time increment $\Delta t$ much longer than the correlation time of the fluctuations but shorter than the resulting diffusion timescale. Note we do not have to say \textit{where} the fluctuations come from, nor do we have to stipulate whether they include a contribution from the perturbed disc's self-gravity.\footnote{The main requirement is just that the stochastic fluctuations are not strong/long-lived enough to trap orbits \`a la \citet{sellwood_binney2002}; see Hamilton, Modak \& Winter-Granic (in preparation) for a detailed discussion.}

Equation~\eqref{eq:QL_diffusion} can be rearranged into the familiar Fokker--Planck form
\begin{equation}
    \label{eq:FP}
    \frac{\partial f_0}{\partial t} = -\sum_i\frac{\partial}{\partial J_i}\l[A_{J_i} f_0
    - \frac{1}{2}\sum_j\frac{\partial}{\partial J_j}\l(D_{J_iJ_j}f_0\r)\r],
\end{equation}
where we have defined the drift coefficients corresponding to the diffusion tensor of Equation~\eqref{eqn:D_JJ} via
\begin{equation}
    \label{eq:drift}
    A_{J_i}(\bJ) = \frac{\overline{\langle\Delta J_i\rangle}}{\Delta t} = \frac{1}{2}\sum_j\frac{\partial D_{J_iJ_j}}{\partial J_j}.
\end{equation}
This drift does not arise from dynamical friction; it is instead an intrinsic consequence of the dependence of the diffusion coefficients on the actions. When diffusion is stronger in some regions of action space than in others, stochastic perturbations can produce systematic changes in the mean actions. The drift coefficients given in Equation~\eqref{eq:drift} account for this effect and must therefore be included in any stochastic realization of Equation~\eqref{eq:QL_diffusion}.

Throughout this paper we take the diffusion tensor to be diagonal, with $D_{J_iJ_j}=0$ for $i\neq j$ (but see \S\ref{sec:beyond}). The drift is then $A_{J_i}=(1/2)\partial D_{J_iJ_i}/\partial J_i$, and the diffusion is set entirely by the coefficients $D_{J_\varphi J_\varphi}$, $D_{J_RJ_R}$ and $D_{J_zJ_z}$.\footnote{\cite{modak_interstellar_2026}, Modak et al. (in preparation), and Hamilton, Modak \& Winter-Granic (in preparation) write these coefficients as $D_{\varphi\varphi}$, $D_{RR}$, and $D_{zz}$. Here we explicitly write out actions in the subscripts: e.g., we write $D_{J_R J_R}$, reserving $D_{RR}$ for the diffusion coefficient of the radius $R$.}

\subsection{Stochastic kick recipe}
\label{sec:kicks}

If we could run our simulations in action space, our problem would now be solved. That is because the Fokker--Planck evolution of $f_0$ described by Equation~\eqref{eq:FP} with a diagonal diffusion tensor is equivalent to the evolution of an ensemble of stars whose actions evolve according to a Langevin equation (e.g., \citealt{gardiner_stochastic_2009}), which we can integrate forward in time by kicking every star according to the Euler--Maruyama update rule
\begin{equation}
    \label{eq:EM_actions}
    J_i \mapsto J_i + A_{J_i}(\bJ)\,\Delta t_\mr{kick} + \eta_i\sqrt{D_{J_iJ_i}(\bJ)\,\Delta t_\mr{kick}}.
\end{equation}
Here, $\Delta t_\mr{kick}$ is the interval between kicks and the $\eta_i$ are independent random numbers, drawn afresh at every step from a normal distribution with zero mean and unit variance, and we evaluate $A_{J_i}$ and $D_{J_iJ_i}$ at the actions at the start of each step. Each star then follows one randomly realized trajectory, and the distribution of infinitely many such trajectories obeys Equation~\eqref{eq:FP} as $\Delta t_\mr{kick}\to 0$.\footnote{To be clear, $\Delta t_\mr{kick}$ is a numerical time-step, not the interval $\Delta t$ of Equation~\eqref{eqn:D_JJ}, and it can be much shorter than the correlation time of the fluctuations. The kicks are uncorrelated from one step to the next, so the correlation of the real fluctuations in time enters only through the coefficients $D_{J_iJ_i}(\bJ)$. The kicked orbits can therefore be expected to reproduce real orbits statistically only on timescales longer than that correlation time, as can Equation~\eqref{eq:QL_diffusion} itself.}

Orbit integrations, however, work in position-velocity space rather than angle-action space. Thus, what we need is a six-dimensional position-velocity kick recipe that reproduces the desired action space diffusion behaviour. If we knew the drift and diffusion of all six angles and actions, those of the positions and velocities would follow from It\^o's lemma, the chain rule for stochastic processes \citep{gardiner_stochastic_2009}. The complication is that in many cases, we have only characterized the three coefficients $D_{J_iJ_i}$. On their own these do not fully specify the kick recipe: they tell us how rapidly a star's actions get kicked on average, but not when those kicks ought to be applied during each orbit, nor how the angles themselves diffuse. In Appendix~\ref{app:prescriptions} we show how a few additional choices allow us to write down a complete recipe. The most important choice is that we assume the statistics of the kicks do not depend on a star's current angles $\bm{\theta}$.
 
Briefly, the recipe is as follows. To reproduce the measured diffusion in $J_R$ ($J_z$) we introduce independent, zero-mean stochastic kicks in $R$ and $v_R$ ($z$ and $v_z$) with angle-independent variances. Diffusion in $J_\varphi$ is different: we implement it by shifting the guiding radius and displacing the star accordingly, leaving $R-\Rg$, $v_R$, $z$ and $v_z$ unchanged. Finally, the kicks must also reproduce the action-space drifts $A_{J_i}$, which we account for by introducing a systematic shift of the guiding radius, and small rescalings of the radial and vertical oscillations. 
%Following the procedure outlined above, we can now express the action-space diffusion and drift coefficients in terms of the position-velocity coordinates $(\varphi,R,z,v_\varphi,v_R,v_z)$. The full derivation is given in Appendix~\ref{app:prescriptions}; here we present the resulting coefficients needed to implement our stochastic kick prescription.
The nonzero diffusion coefficients that result are
%for the radius and azimuthal velocity are
\begin{align}
    \label{eq:DRR_EOM1}
    D_{RR} & =\frac{1}{\Rg^2\Omega^2}\Bigg[\frac{\gamma J_\varphi}{4 J_R}D_{J_RJ_R}
    +\frac{\gamma^4}{4}D_{J_\varphi J_\varphi}\Bigg],\\
    D_{Rv_\varphi} & =-\frac{\gamma}{4J_R}D_{J_RJ_R}
    + \frac{\gamma^2}{2J_\varphi}\l(1-\frac{\gamma^2}{2}\r)D_{J_\varphi J_\varphi},\\
    D_{v_\varphi v_\varphi} & =(\Rg\Omega)^2\l[\frac{\gamma}{4J_R J_\varphi}D_{J_RJ_R}
    +\l(1-\frac{\gamma^2}{2}\r)^2\frac{D_{J_\varphi J_\varphi}}{J_\varphi^2}\r],
%\end{align}
%For the radial velocity and azimuth, we have
%\begin{align}
 \\
    D_{v_Rv_R} & = \frac{(\Rg \Omega)^2 }{\gamma}\frac{D_{J_R J_R}}{J_R J_\varphi},\\
    D_{\varphi v_R} & = \frac{(\Rg \Omega)\gamma}{2J_RJ_\varphi}D_{J_RJ_R},\\
    D_{\varphi\varphi} & = \frac{\gamma^3}{4}\frac{D_{J_R J_R}}{J_R J_\varphi},
%\end{align}
%and for the height and vertical velocity,
%\begin{align}
\\
    D_{zz}&=\frac{1}{2\nu J_z} D_{J_zJ_z},\\
    D_{v_zv_z}&=\frac{\nu}{2J_z} D_{J_zJ_z},
\end{align}
and the corresponding drift coefficients are
\begin{align}
    A_R & = \frac{\gamma^2 \Rg}{2J_\varphi} A_{J_\varphi}
    + (R-\Rg) \l(\frac{A_{J_R}}{2J_R} - \frac{D_{J_R J_R}}{4J_R^2}\r)\notag\\
    &\quad+\frac{\gamma^4 \Rg}{8J_\varphi^2}(2q_R-q-1)D_{J_\varphi J_\varphi},\\
    A_\varphi&=\frac{\gamma v_R}{\Rg\kappa}\l(\frac{A_{J_R}}{2J_R} - \frac{D_{J_RJ_R}}{4J_R^2}\r),\\
    A_{v_R}&=v_R\l(\frac{A_{J_R}}{2J_R} - \frac{D_{J_R J_R}}{4J_R^2}\r),\\
    A_{v_\varphi} & = \frac{A_{J_\varphi}- v_\varphi A_R- D_{R v_\varphi}}{R}, 
    \\
    A_z & = z \left(\frac{A_{J_z}}{2J_z}- \frac{D_{J_zJ_z}}{4J_z^2}\right),\\
    A_{v_z} & = v_z \left(\frac{A_{J_z}}{2J_z}- \frac{D_{J_zJ_z}}{4J_z^2}\right).
    \label{eqn:final_drift}
\end{align}
Here, $q\equiv-\md\ln\Omega/\md\ln R$ and $q_R \equiv -\md\ln\kappa/\md\ln R$ describe the local shear of the circular and epicyclic frequencies, respectively. For reference, $\gamma^2 = 2/(2-q)$, and in a flat rotation curve, we have $q=q_R=1$ and $\gamma = \sqrt{2}$.

Having calculated these coefficients, we can implement them in simulations using Euler–Maruyama update rules. Since some of the position-velocity diffusion coefficients are nonzero off the diagonal, the corresponding stochastic increments must be correlated, so the rules are slightly more complicated than in \eqref{eq:EM_actions}. As shown in Appendix~\ref{app:update}, the resulting prescription requires five independent random numbers $\eta_R$, $\eta_{v_\varphi}$, $\eta_{v_R}$, $\eta_z$ and $\eta_{v_z}$, drawn from a normal distribution with zero mean and unit variance. The update rules are
\begin{align}
    \label{eqn:realspace_update_1}
    R & \mapsto R + A_R\,\Delta t_\mr{kick} + \eta_R \sqrt{D_{RR}\,\Delta t_\mr{kick}}, \\
    v_\varphi & \mapsto v_\varphi + A_{v_\varphi}\Delta t_\mr{kick}
    + \eta_R\,\frac{D_{Rv_\varphi}}{\sqrt{D_{RR}}}\sqrt{\Delta t_\mr{kick}}\notag \\
    &\qquad+ \eta_{v_\varphi}\sqrt{\left(D_{v_\varphi v_\varphi}
    - \frac{D_{Rv_\varphi}^2}{D_{RR}}\right)\Delta t_\mr{kick}}, \\
    v_R & \mapsto v_R + A_{v_R}\Delta t_\mr{kick} + \eta_{v_R}\sqrt{D_{v_R v_R}\,\Delta t_\mr{kick}}, \\
    \varphi & \mapsto \varphi + A_\varphi\,\Delta t_\mr{kick}
    + \eta_{v_R}\,\frac{D_{\varphi v_R}}{\sqrt{D_{v_Rv_R}}}\sqrt{\Delta t_\mr{kick}},\\
    z & \mapsto z + A_z\,\Delta t_\mr{kick} + \eta_{z}\sqrt{D_{zz}\,\Delta t_\mr{kick}},\\
    v_z & \mapsto v_z + A_{v_z}\Delta t_\mr{kick}+\eta_{v_z}\sqrt{D_{v_z v_z}\,\Delta t_\mr{kick}}.
        \label{eqn:realspace_update_6}
\end{align}
The update rules \eqref{eqn:realspace_update_1}--\eqref{eqn:realspace_update_6} (together with the coefficients \eqref{eq:DRR_EOM1}--\eqref{eqn:final_drift}) form the central result of this paper. 
 
\section{Application to ISM-driven diffusion}
\label{sec:ism}
%%%%%%%%%%%%%%%%%%%%%%%%%%%%%%%%%%%%%%%%%%%%%%%%%%%%%%%%%%
%%%%%%%%%%%%%%%%%%%%%%%%%%%%%%%%%%%%%%%%%%%%%%%%%%%%%%%%%%

The recipe of \S\ref{sec:kicks} is valid for epicyclic orbits diffusing according to an {arbitrary} diagonal diffusion tensor $D_{J_i J_i}$, regardless of its functional form or the physical origin of the perturbations. Here, we apply the recipe to the particular case of ISM-driven diffusion. We begin in \S\ref{sec:ism_coefficients} by summarizing the diffusion coefficients measured by \citet{modak_interstellar_2026} and Modak et al. (in preparation). Then in \S\ref{sec:box} we perform two tests of the recipe: one in the same local 3D shearing box that \citet{modak_interstellar_2026} and Modak et al. (in preparation) used for their simulations, and another in a global Milky Way potential.

%%%%%%%%%%%%%%%%%%%%%%%%%%%%%%%%%%%%%%%%%%%%%%%%%%%%%%%%%%
%%%%%%%%%%%%%%%%%%%%%%%%%%%%%%%%%%%%%%%%%%%%%%%%%%%%%%%%%%
\subsection{Diffusion coefficients}
\label{sec:ism_coefficients}
%%%%%%%%%%%%%%%%%%%%%%%%%%%%%%%%%%%%%%%%%%%%%%%%%%%%%%%%%%
%%%%%%%%%%%%%%%%%%%%%%%%%%%%%%%%%%%%%%%%%%%%%%%%%%%%%%%%%%

\citet{modak_interstellar_2026} integrated the orbits of test stars through the time-dependent gravitational potential of the ISM in the TIGRESS-NCR simulation of a patch of the Galactic disc at the solar radius (the `R8' model of \citealt{Kim2023}). They measured the resulting radial and azimuthal diffusion behaviour of the test stars, and thereby read off the following (radial action-dependent) diffusion coefficients for the solar neighbourhood:
\begin{equation}
    \label{eq:Dpp_R8}
    D^{\odot}_{J_\varphi J_\varphi} \simeq
    \begin{cases}
        6900\,\dfrac{\mr{(kpc\ km/s)^2}}{\mr{Gyr}} & (J_R\ll J_{\perp,*}), \\[2.5ex]
        5900\,\dfrac{\mr{(kpc\ km/s)^2}}{\mr{Gyr}}\times \l(\dfrac{J_R}{\mr{kpc\ km/s}}\r)^{-1/2} & (J_R\gg J_{\perp,*}),
    \end{cases}
\end{equation}
\begin{equation}
    \label{eq:DRR_R8}
    D^{\odot}_{J_RJ_R} \simeq
    \begin{cases}
        9.3\,\dfrac{\mr{(kpc\ km/s)^2}}{\mr{Gyr}}\times \l(\dfrac{J_R}{\mr{kpc\ km/s}}\r) & (J_R \ll J_{\perp,*}),\\[2.5ex]
        35\,\dfrac{\mr{(kpc\ km/s)^2}}{\mr{Gyr}}\times \l(\dfrac{J_R}{\mr{kpc\ km/s}}\r)^{-1/2} & (J_R \gg J_{\perp,*}).
    \end{cases}
\end{equation}
Similarly, Modak et al. (in preparation) characterized the resulting vertical action-space diffusion in the same ISM potential, measuring a solar-neighbourhood vertical diffusion coefficient of 
\begin{equation}
    \label{eq:Dzz_R8}
    D^{\odot}_{J_zJ_z}\simeq
    \begin{cases}
        0.25\,\dfrac{\mr{(kpc\ km/s)^2}}{\mr{Gyr}}
        \l(\dfrac{J_z}{\mr{kpc\ km/s}}\r)
        \\[1ex]
        \hfill (J_z \ll J_{z,*},\quad J_R \ll J_{\perp,*}), \\[2.5ex]
        0.30\,\dfrac{\mr{(kpc\ km/s)^2}}{\mr{Gyr}}
        \l(\dfrac{J_z}{\mr{kpc\ km/s}}\r)
        \l(\dfrac{J_R}{\mr{kpc\ km/s}}\r)^{-1/2}
        \\[1ex]
        \hfill (J_z \ll J_{z,*},\quad J_R \gg J_{\perp,*}), \\[2.5ex]
        0.31\,\dfrac{\mr{(kpc\ km/s)^2}}{\mr{Gyr}}
        \l(\dfrac{J_z}{\mr{kpc\ km/s}}\r)^{-1/2}
        \l(\dfrac{J_R}{\mr{kpc\ km/s}}\r)^{-1/2}
        \\[1ex]
        \hfill (J_z \gg J_{z,*},\quad J_R \gg J_{\perp,*}).
    \end{cases}
\end{equation}
The two branches of the radial and azimuthal coefficients correspond to different dynamical regimes: in the \textit{radial long-wavelength} regime ($J_R < J_{\perp,*}$), the radial epicycle of a star is smaller than the dominant wavelength of the ISM fluctuations, $\lambda_*\simeq600$\,pc; in the \textit{radial short-wavelength} regime ($J_R > J_{\perp,*}$) it is larger. The transition action formally separating the asymptotic regimes, $J_{\perp,*} \equiv \kappa\lambda_*^2/(8\pi^2)\simeq 0.2$\,kpc\,km\,s$^{-1}$, is the value of $J_R$ at which $k_*a_R=1$, where $k_*=2\pi/\lambda_*$. However, for the diffusion coefficients to be continuous in this ``broken power-law'' model, we require the transition between the branches to occur where they cross, at $J_R\simeq0.73$\,kpc\,km\,s$^{-1}$ for $D_{J_\varphi J_\varphi}$ and $J_R\simeq 2.4$\,kpc\,km\,s$^{-1}$ for $D_{J_RJ_R}$. Similarly, the vertical coefficient above has \textit{three} branches, since we have to additionally stipulate whether the vertical action $J_z$ is much smaller or much larger than the critical value $J_{z,*} \simeq 0.09 $\,kpc\,km\,s$^{-1}$ (Modak et al., in preparation), corresponding to \textit{vertical long-} and \textit{vertical short-wavelength} regimes respectively.\footnote{In realistic Galactic environments, stellar orbits heat radially much more rapidly than they do vertically (Modak et al., in preparation), so the long radial wavelength, short vertical wavelength regime rarely occurs in practice. In our diffusion coefficient implementation, stars in this regime continue to diffuse with the long radial wavelength, long vertical wavelength diffusion coefficient. As a result $D_{J_zJ_z}$ is discontinuous across $J_R\simeq1.4$\,kpc\,km\,s$^{-1}$ for $J_z\gtrsim1$\,kpc\,km\,s$^{-1}$. We find that this makes negligible difference for our results.} %about 3 per cent of the stars of \S\ref{sec:box} are in this regime after 10\,Gyr.}
As with the planar coefficients, we impose continuity on the vertical diffusion coefficients, which sets the transitions to the radial and vertical short-wavelength regimes at $J_R\simeq 1.4$\,kpc\,km\,s$^{-1}$ and $J_z\simeq 1.0$\,kpc\,km\,s$^{-1}$, respectively.

These coefficients apply only to ISM-driven diffusion at the solar radius, but \citet{modak_interstellar_2026} also derived how the radial and azimuthal coefficients scale with the properties of the gas and of the Galaxy. Briefly, for non-solar neighbourhood conditions one replaces the above $D_{J_R J_R}^{\odot}$ and $D_{J_\varphi J_\varphi}^{\odot}$ by the scaled quantities
\begin{align}
    \label{eq:Dpp_long}
    D_{J_\varphi J_\varphi}(\bJ) &\simeq \l(\frac{\overline{\Sigma}(\Rg)}{12\,\Msun\,\mathrm{pc}^{-2}}\r)^2 \l(\frac{\lambda_*}{600\,\mathrm{pc}}\r)^{2.3}\l(\frac{\tau_*}{70\,\mathrm{Myr}}\r) \nn\\
    & \quad\quad\quad\quad\quad\quad\times\l(\frac{\Rg}{8\,\mathrm{kpc}}\r)^{2} \l(\frac{\xi(J_z, \nu, h)}{0.98}\r)D_{J_\varphi J_\varphi}^{\odot}(J_R)
\end{align}
\begin{align}
    \label{eq:DRR_long}
    D_{J_RJ_R}(\bJ) &\simeq \l(\frac{\overline{\Sigma}(\Rg)}{12\,\Msun\,\mathrm{pc}^{-2}}\r)^2 \l(\frac{\lambda_*}{600\,\mathrm{pc}}\r)^{2.3}\l(\frac{\tau_*}{70\,\mathrm{Myr}}\r) \nn\\
    & \quad\times\l(\frac{\kappa}{40\,\mathrm{km\,s^{-1}\,kpc^{-1}}}\r)^{-1} \l(\frac{\xi(J_z, \nu, h)}{0.94}\r)D_{J_RJ_R}^{\odot}(J_R)
\end{align}
in the long-wavelength regime, and
\begin{align}
    \label{eq:Dpp_short}
    D_{J_\varphi J_\varphi}(\bJ) &\simeq \l(\frac{\overline{\Sigma}(\Rg)}{12\,\Msun\,\mathrm{pc}^{-2}}\r)^2 \l(\frac{\lambda_*}{600\,\mathrm{pc}}\r)^{3.3}\l(\frac{\tau_*}{70\,\mathrm{Myr}}\r) \l(\frac{\Rg}{8\,\mathrm{kpc}}\r)^{2} \nn\\
    &\quad\times \l(\frac{\kappa}{40\,\mathrm{km\,s^{-1}\,kpc^{-1}}}\r)^{1/2} \l(\frac{\xi(J_z, \nu, h)}{0.84}\r)D_{J_\varphi J_\varphi}^{\odot}(J_R)
\end{align}
\begin{align}
    \label{eq:DRR_short}
    D_{J_RJ_R}(\bJ) &\simeq \l(\frac{\overline{\Sigma}(\Rg)}{12\,\Msun\,\mathrm{pc}^{-2}}\r)^2
    \l(\frac{\lambda_*}{600\,\mathrm{pc}}\r)^{5.3}\l(\frac{\tau_*}{70\,\mathrm{Myr}}\r) \nn\\
    &\quad\times\l(\frac{\kappa}{40\,\mathrm{km\,s^{-1}\,kpc^{-1}}}\r)^{1/2}
    \l(\frac{\xi(J_z, \nu, h)}{0.75}\r)D_{J_R J_R}^{\odot}(J_R)
\end{align}
in the short-wavelength regime. For the vertical diffusion coefficient, Modak et al. (in preparation) similarly derived
\begin{align}
    \label{eq:Dzz_longlong}
    D_{J_z J_z}(\bJ) & \simeq \l(\frac{\overline{\Sigma}(\Rg)}{12\,\Msun\,\mathrm{pc}^{-2}}\r)^2 \l(\frac{\lambda_*}{600\,\mathrm{pc}}\r)^{2.3}\l(\frac{\tau_*}{70\,\mathrm{Myr}}\r) \nn\\
    & \times \l(\frac{1 + 2\pi h/\lambda_*}{3.1}\r)^{-2} \l(\frac{\nu}{81\,\mathrm{km\,s^{-1}\,kpc^{-1}}}\r)^{-1}  D_{J_z J_z}^{\odot}(J_R, J_z)
\end{align}
in the vertical long-wavelength, radial long-wavelength regime,
\begin{align}
    \label{eq:Dzz_longshort}
    D_{J_z J_z}(\bJ) & \simeq \l(\frac{\overline{\Sigma}(\Rg)}{12\,\Msun\,\mathrm{pc}^{-2}}\r)^2 \l(\frac{\lambda_*}{600\,\mathrm{pc}}\r)^{3.3}\l(\frac{\tau_*}{70\,\mathrm{Myr}}\r) \nn \\
    & \quad \times \l(\frac{1 + 2\pi h/\lambda_*}{3.1}\r)^{-2} \l(\frac{\nu}{81\,\mathrm{km\,s^{-1}\,kpc^{-1}}}\r)^{-1} \nn \\
    & \quad\quad \times \l(\frac{\kappa}{40\,\mathrm{km\,s^{-1}\,kpc^{-1}}}\r)^{1/2} D_{J_z J_z}^{\odot}(J_R, J_z)
\end{align}
in the vertical long-wavelength, radial short-wavelength regime, and
\begin{align}
    \label{eq:Dzz_shortshort}
    D_{J_z J_z}(\bJ) & \simeq \l(\frac{\overline{\Sigma}(\Rg)}{12\,\Msun\,\mathrm{pc}^{-2}}\r)^2 \l(\frac{\lambda_*}{600\,\mathrm{pc}}\r)^{6.3}\l(\frac{\tau_*}{70\,\mathrm{Myr}}\r) \nn \\
    & \quad \times \l(\frac{1 + 2\pi h/\lambda_*}{3.1}\r)^{-2} \l(\frac{\nu}{81\,\mathrm{km\,s^{-1}\,kpc^{-1}}}\r)^{1/2} \nn \\
    & \quad\quad \times \l(\frac{\kappa}{40\,\mathrm{km\,s^{-1}\,kpc^{-1}}}\r)^{1/2} D_{J_z J_z}^{\odot}(J_R, J_z)
\end{align}
in the vertical short-wavelength, radial short-wavelength regime. Here, $\overline{\Sigma}$ is the mean surface density of the gas evaluated at the guiding centre, $\lambda_*$ is the dominant substructure spatial scale driving the diffusion, $\tau_*$ is the correlation time of fluctuations at scale $\lambda_*$, and $\xi\le1$ is the factor by which vertical motion suppresses radial transport, which depends on vertical action, the vertical frequency, and the thickness of the gas layer $h$---see Equation (48) of \citet{modak_interstellar_2026} for an approximate fitting formula for it. The numbers that divide $\xi$ in Equations~\eqref{eq:Dpp_long}--\eqref{eq:DRR_short} are its measured values for the simulated stars in each regime, so that the scaled coefficients reduce to Equations~\eqref{eq:Dpp_R8} and \eqref{eq:DRR_R8} at the solar radius. We also remind the reader that the transition radial and vertical action values separating the broken power laws must be updated accordingly to ensure continuity of the scaled coefficients.

We have implemented the coefficients given in Equations \eqref{eq:Dpp_long}--\eqref{eq:Dzz_shortshort} into the \texttt{gala} code \citep[v1.13;][]{pricewhelan2017}; the numerical values characterizing the fluctuation strengths and scales can easily be tweaked to account for different galactic/ISM conditions. For reference, integrating 50\,000 test particle orbits for 10\,Gyr through \texttt{gala}'s \texttt{MilkyWayPotential2022} potential while following the recipe presented in \S\ref{sec:kicks} to deliver diffusive kicks every $\Delta t_{\mathrm{kick}} = 1\,$Myr on a single core of an Intel Xeon Gold 6248R processor 
(3.00\,GHz) takes $\sim 0.16\,$CPU-hours, compared to $\sim 0.04\,$CPU-hours for orbit integration without any kicks, while the same number of orbits directly integrated through the TIGRESS-NCR fluctuations (including time required to read in simulation snapshots and evaluate the kicks) takes $\sim 8\,$CPU-hours.

%%%%%%%%%%%%%%%%%%%%%%%%%%%%%%%%%%%%%%%%%%%%%%%%%%%%%%%%%%
%%%%%%%%%%%%%%%%%%%%%%%%%%%%%%%%%%%%%%%%%%%%%%%%%%%%%%%%%%
\subsection{Tests in a shearing box and in a global potential}
\label{sec:box}
%%%%%%%%%%%%%%%%%%%%%%%%%%%%%%%%%%%%%%%%%%%%%%%%%%%%%%%%%%
%%%%%%%%%%%%%%%%%%%%%%%%%%%%%%%%%%%%%%%%%%%%%%%%%%%%%%%%%%

As a test of the recipe we first integrated test-particle trajectories in the same 3D periodic shearing box that \citet{modak_interstellar_2026} used. The shearing box is centred at $R_0=8$\,kpc and assumes a locally flat rotation curve, with $\Omega=28$\,km\,s$^{-1}$\,kpc$^{-1}$ and a vertical epicyclic period $2\pi/\nu \approx 76$\,Myr. As in \citet{modak_interstellar_2026}, we initialized 50\,000 stars on circular orbits at random positions in the midplane ($z=0$) of this box and integrated their orbits for 10\,Gyr. Every $\Delta t_\mathrm{kick}=1$\,Myr we kicked each star according to the recipe of \S\ref{sec:framework} and Appendix~\ref{app:prescriptions}, using the local ISM diffusion coefficients \eqref{eq:Dpp_R8}--\eqref{eq:Dzz_R8}. Between kicks, we advanced each star in the plane with the exact solution of the shearing-box equations of motion, and in the vertical direction in the stellar disk and dark matter potential of the R8 model (Equation (16) of \citealt{modak_interstellar_2026}, to which we added a mean gas layer of surface density $12\,\Msun\,\mathrm{pc}^{-2}$ whose thickness we chose to be $200\,$pc to match the vertical period quoted above.

We then repeated the test in a global potential, the \texttt{MilkyWayPotential2022} of \texttt{gala} \citep{pricewhelan2017}. We initialized 50\,000 stars on circular orbits in the midplane ($z = 0$), at radii between 7.5 and 8.5\,kpc, integrated their orbits for 10\,Gyr, and kicked them every 1\,Myr in the same way, with the diffusion coefficients built into \texttt{gala}. To be clear, this second test is not a simulation of an entire Galactic disc---its purpose is to find out whether the effects neglected by the shearing box, such as the curvature of the global coordinate system, the variation of the orbital frequencies and of the diffusion coefficients with galactocentric radius, and the deviation of the orbit from the epicyclic approximation, alter the results.

\begin{figure}
    \centering
    \includegraphics[width=\columnwidth]{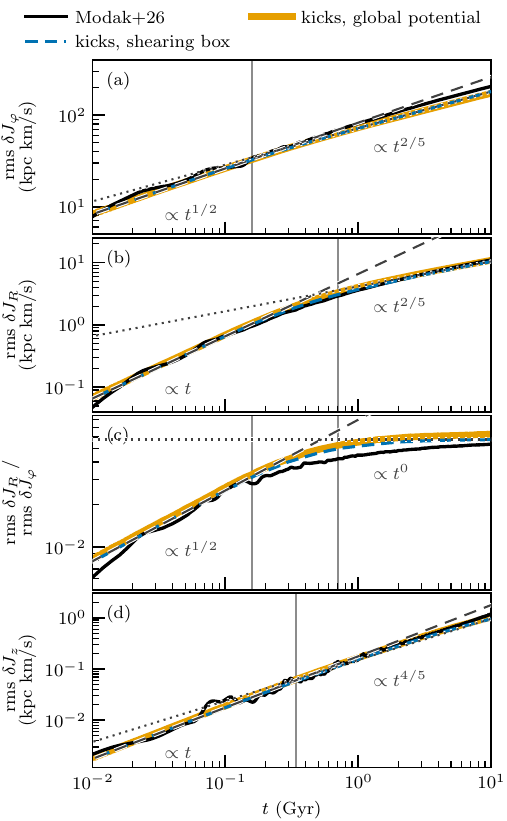}
    \caption{Heating and migration in action space for the orbit integrations through the simulated TIGRESS-NCR fluctuations (black), stars kicked according to the recipe presented in this work in the shearing box (blue dashed), and stars kicked according to the same recipe in the global \texttt{MilkyWayPotential2022} potential (gold). Panels (a), (b) and (d) show the rms changes (equation~\ref{eq:rms}) in the azimuthal, radial and vertical actions of 50\,000 stars that start on circular orbits in the midplane. Panel (c) shows the radial heating-to-migration ratio. Dashed and dotted grey lines show the power laws in time expected in the radial long-wavelength and short-wavelength regimes. Vertical grey lines show the time of transition between the two regimes, defined as the time at which the mean radial action of the stars kicked in the shearing box reaches the action at which the two branches of the diffusion coefficient for that panel cross (for $D_{J_zJ_z}$ in panel (d), its first two branches, at $J_R\simeq1.4$\,kpc\,km\,s$^{-1}$). On panel (c), we show the transition times of both panels (a) and (b).}
    \label{fig:box_actions}
\end{figure}

\begin{figure*}
    \centering
    \includegraphics[width=0.98\linewidth]{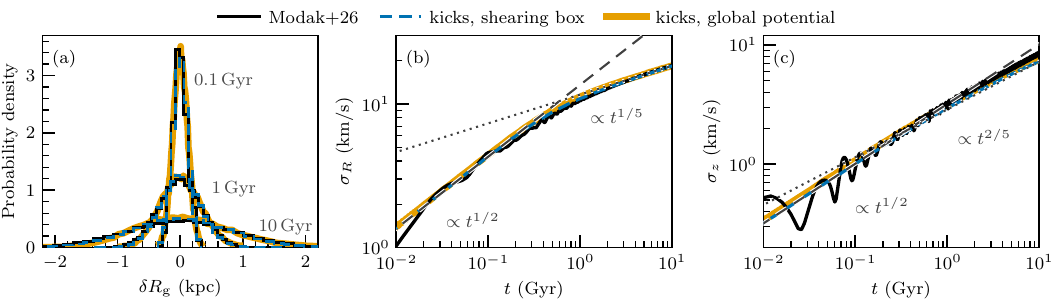}
    \caption{Changes in guiding radius (panel (a)) and radial/vertical velocity dispersions (panels (b) and (c) respectively) for the stars of Figure~\ref{fig:box_actions}. Panel (a) shows the distribution of the change in guiding radius $\delta\Rg$ at 0.1, 1 and 10\,Gyr. Colours and grey lines are as in Figure~\ref{fig:box_actions}.}
    \label{fig:box_observables}
\end{figure*}

\begin{figure}
    \centering
    \includegraphics[width=0.98\linewidth]{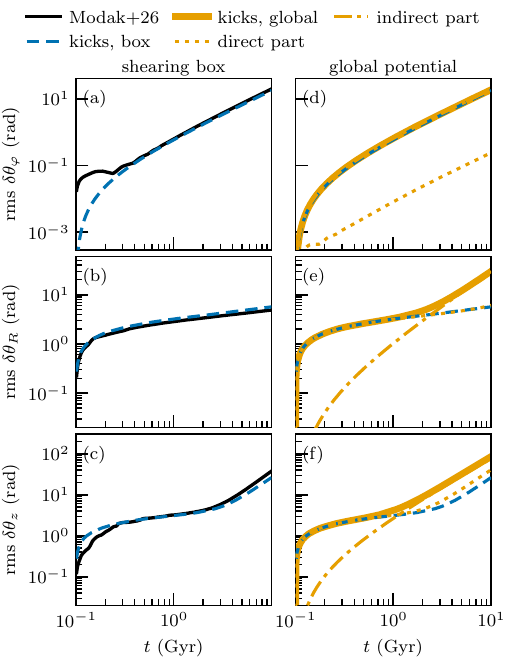}
    \caption{Diffusion of the angles for the stars of Figure~\ref{fig:box_actions}. From top to bottom, the rows show the rms changes in $\theta_\varphi$, $\theta_R$ and $\theta_z$ relative to unperturbed orbits, measured from an initial time of $t_0=100$\,Myr. The left column shows the shearing box. Black lines show the orbits integrated through TIGRESS-NCR by \citet{modak_interstellar_2026}, and dashed blue lines show stars kicked in the shearing box. The right column shows the global potential. Solid gold lines show the total change for stars kicked in the global potential, dotted and dash-dotted gold lines show its direct and indirect parts, and dashed blue lines show the total change in the shearing box.}
    \label{fig:box_angles}
\end{figure}

In Figures~\ref{fig:box_actions}--\ref{fig:box_angles} we show the results of both tests, compared with the same measurements taken from the simulations of \cite{modak_interstellar_2026} and Modak et al. (in preparation). In particular we measure the change $\delta X$ in a quantity $X$ of the stars by its root-mean-square value over the $N=50\,000$ stars,
\begin{equation}
    \label{eq:rms}
    \mathrm{rms}\,\delta X(t) \equiv \l[\frac{1}{N}\sum_{n=1}^{N}\delta X_n(t)^2\r]^{1/2},
\end{equation}
where $\delta X_n(t) \equiv X_n(t) - X_n(t_0)$ is the change in $X$ for star $n$ between time $t$ and some initial time $t_0$. For the actions and the guiding radius, we take $t_0 = 0$; for the angles, we use the change defined in Equation~\eqref{eq:dtheta} below, measured from a later time $t_0$.

First, in Figure~\ref{fig:box_actions} we plot the rms changes in the three actions, and the heating-to-migration ratio rms\,$\delta J_R/$rms\,$\delta J_\varphi$, over the 10\,Gyr of the simulation. The vertical grey line in each panel marks the time of transition from the long radial wavelength to the short radial wavelength regime, i.e., the time at which the mean radial action of the kicked stars reaches the action at which the two branches of the relevant diffusion coefficient cross (see the discussion below Equation \eqref{eq:Dzz_R8}). These times lie between 0.16 and 0.70\,Gyr, and each curve bends from one power law to the other around its own transition time. The agreement between our kick procedure and the direct orbit integrations is very good, differing at most at the $\sim 10\%$ level. This is very encouraging given how computationally cheap our approach is, and the crudeness of the transition between dynamical regimes built into our coefficients \eqref{eq:Dpp_R8}--\eqref{eq:Dzz_R8}.\footnote{Preliminary tests suggest that a smoother transition between regimes would give rise to even better agreement, although we have not explored this in detail.} Also, we see that there is almost no difference between the results of the shearing box test and the corresponding global run. The effects that the shearing box leaves out therefore appear to matter little for ISM-driven heating and migration.

Secondly, in Figure~\ref{fig:box_observables} we plot the same comparison for quantities that are closer to observations, namely the distribution of the change in guiding radius $\delta\Rg$ at three times, and the radial and vertical velocity dispersions $\sigma_R$ and $\sigma_z$. Once again the agreement is excellent. We take this opportunity to re-emphasize two lessons from \cite{modak_interstellar_2026}, now confirmed in the global context: (i) ISM-driven radial migration is significant, with about a quarter of the stars in the solar neighbourhood migrating $\gtrsim 1$\,kpc over $10$\,Gyr, and (ii) the transition to the short-wavelength regime drastically suppresses radial heating, from a random walk $\sigma_R \propto t^{1/2}$ that would be expected for temporally uncorrelated kicks to a subdiffusive $\sigma_R \propto t^{1/5}$ that arises because most radial forcing cancels out across an epicycle.

Thirdly, we have not put any information about angle space diffusion into our recipe, except for the assumption of \S\ref{sec:framework} that the statistics of the kicks which a star receives do not depend on $\btheta$. It is therefore important to check that our successful recovery of the direct orbit integrations' action-space results does not also introduce spurious behaviour in the angles. To do this, we measured the change in each angle relative to an unperturbed orbit,
\begin{equation}
    \label{eq:dtheta}
    \delta\theta_i(t) \equiv \theta_i(t)-\theta_i(t_0)-\Omega_{i}\,(t-t_0),
\end{equation}
where $i=\varphi,R,z$ and $\Omega_i=(\Omega,\kappa,\nu)$ are the frequencies of the star's orbit at the reference time $t_0$. We do not measure the changes in the angles from $t=0$, as we do for the actions, because the stars start on circular orbits in the midplane, where $\theta_R$ and $\theta_z$ are not well-defined. We therefore measure them from a reference time $t_0=100$\,Myr, which is somewhat longer than the correlation time $\tau_*\simeq70$\,Myr of the fluctuations and allows finite radial and vertical oscillation amplitudes to develop. We have verified that our results are insensitive to the choice of $t_0$ for $t_0\geq100$\,Myr.

In the left column of Figure~\ref{fig:box_angles} we plot rms\,$\delta\theta_i$, defined as in Equation~\eqref{eq:rms}, for the orbits kicked in the shearing box and for the orbits of \citet{modak_interstellar_2026}. The agreement across all three angles is at the level of $\sim 10\%$, and often much better. The most notable exception is the discrepancy in $\delta\theta_\varphi$ at very early times (panel a), which arises because our recipe does not capture `direct' kicks to $\theta_\varphi$ (see Equation \eqref{eq:dtheta_ind} below). However, these become unimportant once the changes in orbital frequency driven by radial migration dominate the evolution of $\delta\theta_\varphi$. 
%We put no measured information about the angles into the kick recipe. All that it contains about them is the assumption of \S\ref{sec:framework} that the statistics of the kicks which a star receives do not depend on its angles $\btheta$.
Relatedly, in the right column of Figure~\ref{fig:box_angles} we compare the shearing box results (blue dashed lines) with those from the global potential. For $\theta_\varphi$ the two agree at the $\sim 10\%$ level at all times. For $\theta_R$ and $\theta_z$, while they agree at $t\lesssim 1$\,Gyr, they do \textit{not} agree at later times. This is because in the global potential the frequencies $\kappa$ and $\nu$ depend on the star's guiding radius, so they change as the star migrates, and the angles drift differently as a result. In other words, much of the diffusion in $\bm{\theta}$ arising in the global simulation is `indirect', driven by action space kicks that change stars' frequencies, rather than angle space kicks directly. To separate these two effects, we write $\delta\theta_i=\delta\theta_{i,\mathrm{ind}}+\delta\theta_{i,\mathrm{dir}}$ where
\begin{equation}
    \label{eq:dtheta_ind}
    \delta\theta_{i,\mathrm{ind}}(t)=\int_{t_0}^t\l[\Omega_i(t')-\Omega_i\r]\md t',
\end{equation}
where $\Omega_i(t')$ is the frequency evaluated at the guiding radius of the star at time $t'$; the \textit{direct} part $\delta\theta_{i,\mathrm{dir}}$ is defined to be the difference between the full $\delta\theta_i$ and the indirect contribution. We see from the gold dotted and dash-dotted lines in Figure \ref{fig:box_angles} that the indirect part always dominates at late times.
%\footnote{For $\theta_\varphi$ (panel (d)), the indirect part dominates always in both shearing box and global tests, because the kicks do not change $\theta_\varphi$ directly at all (\S\ref{sec:framework}).} 
This gives us further confidence that supplying correct action space diffusion is what really matters, and that our assumption of $\theta$-independent kick variances, which affects only the direct part, matters little at late times in a global potential.

We perform further checks of our recipe and some convergence tests in Appendix \ref{app:convergence}.

\section{Discussion}
\label{sec:discussion}
%%%%%%%%%%%%%%%%%%%%%%%%%%%%%%%%%%%%%%%%%%%%%%%%%%%%%%%%%%
%%%%%%%%%%%%%%%%%%%%%%%%%%%%%%%%%%%%%%%%%%%%%%%%%%%%%%%%%%

In this section we first show why a recipe for ISM-driven diffusion has to kick the positions of stars as well as their velocities (\S\ref{subsec:velocity_kicks}). We then apply the recipe to two simple models of phase space structures in the Milky Way's disc (\S\ref{sec:illustrations}), and finally discuss its limitations and possible extensions (\S\ref{sec:beyond}).
%%%%%%%%%%%%%%%%%%%%%%%%%%%%%%%%%%%%%%%%%%%%%%%%%%%%%%%%%%
%%%%%%%%%%%%%%%%%%%%%%%%%%%%%%%%%%%%%%%%%%%%%%%%%%%%%%%%%%
\subsection{Kicks in velocity only}
\label{subsec:velocity_kicks}
%%%%%%%%%%%%%%%%%%%%%%%%%%%%%%%%%%%%%%%%%%%%%%%%%%%%%%%%%%
%%%%%%%%%%%%%%%%%%%%%%%%%%%%%%%%%%%%%%%%%%%%%%%%%%%%%%%%%%

Our recipe kicks the positions of stars as well as their velocities. Gravitational forces, however, change only velocities, so the kicks in position may sound unphysical, and a simpler recipe would kick the velocities alone. Could such a recipe reproduce a given diffusion in action space?

For the ISM, the answer is no. To see why, let us ignore vertical motion for simplicity. Suppose that we kick impulsively in $v_\varphi$ and $v_R$, and that the statistics of these kicks do not depend on angles, as in our recipe. In this subsection only, let $D_{v_\varphi v_\varphi}$ and $D_{v_Rv_R}$ denote the diffusion coefficients of these kicks in velocity, and not those of our recipe in \S\ref{sec:kicks}. Since $J_\varphi = Rv_\varphi$, a kick changes the angular momentum by $\Delta J_\varphi = R\,\Delta v_\varphi$. We calculate the resulting $D_{J_\varphi J_\varphi}$ by squaring this and ensemble-averaging, which in this case means averaging over many weak kicks at a fixed spatial position $\bm{r}(\btheta,\bJ)$, and then averaging over angles $\btheta$ at fixed actions $\bJ$. Using the fact that the angle-average of $R^2$ returns $\Rg^2(J_\varphi)$ up to corrections of second order in epicyclic amplitude, we find to this accuracy
\begin{equation}
    D_{J_\varphi J_\varphi} = \frac{J_\varphi}{\Omega}\, D_{v_\varphi v_\varphi}.
    \label{eqn:D_phiphi_Formal}
\end{equation}
To get the corresponding result for $D_{J_RJ_R}$, we calculate the change in $J_R = (1/2)v_R^2/\kappa + (1/2)\kappa (R-\Rg)^2$ due to the kick, square it, and perform the same averages. We find, to the same accuracy,
\begin{equation}
    D_{J_RJ_R} = \frac{J_R}{\kappa}\left(\gamma^2 D_{v_\varphi v_\varphi} + D_{v_Rv_R}\right).
    \label{eqn:D_RR_Formal}
\end{equation}
The term in $D_{v_\varphi v_\varphi}$ arises because a kick in $v_\varphi$ moves the guiding radius but leaves the star where it was, and so changes $R-\Rg$. It follows from equations~\eqref{eqn:D_phiphi_Formal} and~\eqref{eqn:D_RR_Formal} that
\begin{equation}
    D_{v_Rv_R} = \frac{\kappa}{J_R}\,D_{J_RJ_R} - \gamma^2\,\frac{D_{J_\varphi J_\varphi}}{\Rg^2}.
    \label{eq:DvRvR_vel}
\end{equation}
This coefficient is negative, and hence unphysical, whenever
\begin{equation}
    J_R > \frac{2}{\gamma^3}\,\frac{D_{J_RJ_R}}{D_{J_\varphi J_\varphi}}\,J_\varphi.
    \label{eq:JR_negativity_condition}
\end{equation}
For the short-wavelength coefficients of \S\ref{sec:ism_coefficients}, this criterion is  $J_R > 7.5$\,kpc\,km\,s$^{-1}$ at the solar radius, whatever the amplitude of the ISM fluctuations. This is the mean radial action of a population of stars with a radial velocity dispersion $\sigma_R \simeq 17$\,km\,s$^{-1}$. Pure velocity kicks are therefore not capable of producing the correct diffusive behaviour for such populations. Physically, above this threshold the kicks in $v_\varphi$ needed for a given rate of migration heat the stars faster than $D_{J_RJ_R}$ allows, even with no kicks in $v_R$.

One might try to circumvent this problem by simply setting $D_{v_Rv_R}=0$ wherever equation~\eqref{eq:DvRvR_vel} is negative, and adding drifts that give the correct mean change in each action. However, such kicks still fail above the threshold \eqref{eq:JR_negativity_condition}. A drift changes the mean of each $\Delta J_R$ but not its variance, and it is the variance per unit time that defines $D_{J_RJ_R}$. Above the threshold the kicks in $v_\varphi$ alone change $J_R$ with a variance per unit time $\gamma^2(J_R/\kappa)D_{v_\varphi v_\varphi}$, which is larger than $D_{J_RJ_R}$. Kicks in velocity can therefore give the correct $D_{J_\varphi J_\varphi}$ or the correct $D_{J_RJ_R}$, but not both. Our position-velocity kick recipe has no such issue, because a star that is kicked in $J_\varphi$ is moved together with its guiding centre, so that its $J_R$ does not change. The three kinds of kick of \S\ref{sec:kicks} therefore change the three actions independently, and they reproduce the diffusion equation of \S\ref{sec:action_diffusion} for any non-negative coefficients $D_{J_iJ_i}$, to the first order in epicyclic amplitude at which we work. This is something pure velocity kicks cannot achieve.

This failure agrees with the finding of \citet{modak_interstellar_2026} that fluctuations with the spatial structure of the ISM but no correlation in time drive too much heating for a given amount of migration. Independent kicks in velocity, with statistics that do not depend on the angles, represent such fluctuations. The kicks in position allow our recipe, which is also a series of independent kicks, to reproduce the effect of fluctuations that are correlated in time.

One might still have philosophical objections to the very idea of position-space kicks in a calculation that is supposed to simulate gravitational forces. To this we offer the counterargument that $N$-body (or test-particle) galaxy simulations are \textit{not} designed to simulate the motion of $N$ individual stars. Instead, a simulation is just a Monte-Carlo realization of a solution to the Boltzmann equation describing the probability density function of $\gg N$ stars (and Equation \eqref{eq:FP} is an approximate version of that Boltzmann equation). Thus, $N$-body simulations are only ever valid statistically \citep{quinlan1992reliability}, and are only meant to tell us how discrete blobs of probability fluid (each representing the average density of many stars) move through phase space. If our recipe solves the Boltzmann equation properly, it is doing what such simulations were designed for.

%%%%%%%%%%%%%%%%%%%%%%%%%%%%%%%%%%%%%%%%%%%%%%%%%%%%%%%%%%
%%%%%%%%%%%%%%%%%%%%%%%%%%%%%%%%%%%%%%%%%%%%%%%%%%%%%%%%%%
\subsection{Two illustrations in the Milky Way}
\label{sec:illustrations}
%%%%%%%%%%%%%%%%%%%%%%%%%%%%%%%%%%%%%%%%%%%%%%%%%%%%%%%%%%
%%%%%%%%%%%%%%%%%%%%%%%%%%%%%%%%%%%%%%%%%%%%%%%%%%%%%%%%%%

What does ISM-driven diffusion do to the fine-grained structures that \textit{Gaia} has found, or is predicted to find, in the Galactic disc? Here, we illustrate its effect on two of them: the tree-ring structure of stars trapped at the corotation resonance of the Galactic bar \citep{chiba_tree-ring_2021}, and the spiral in the vertical phase space $(z,v_z)$ of stars near the Sun \citep{antoja2018,binney_origin_2018}. We intend these examples as illustrations of the recipe and the qualitative effects of diffusion driven by a realistic ISM rather than as fits to the data. In each case, we evolve copies of the same initial stellar population containing 10 million stars under three different diffusion prescriptions. The first has no diffusion, and the second has the ISM-driven diffusion of \S\ref{sec:ism}. In the third, the long-wavelength coefficients of \S\ref{sec:ism_coefficients} hold at all $J_R$ and $J_z$, which is what a series of impulsive velocity kicks that are uncorrelated in time would produce (\citealt{modak_interstellar_2026}; Hamilton, Modak, \& Winter-Granic, in preparation).\footnote{To be clear, the third integration is not the GMC-driven scattering of \citet{spitzerschwarzchild_51,spitzerschwarzchild_53}. In that picture the heating rate falls as the random velocity $v$ of a star grows (leading to, e.g., $\sigma_R\propto t^{1/3}$), whereas in our temporally uncorrelated runs it is independent of $v$ (leading to $\sigma_R\propto t^{1/2}$).}

\begin{figure*}
    \centering
    \includegraphics[width=0.82\linewidth]{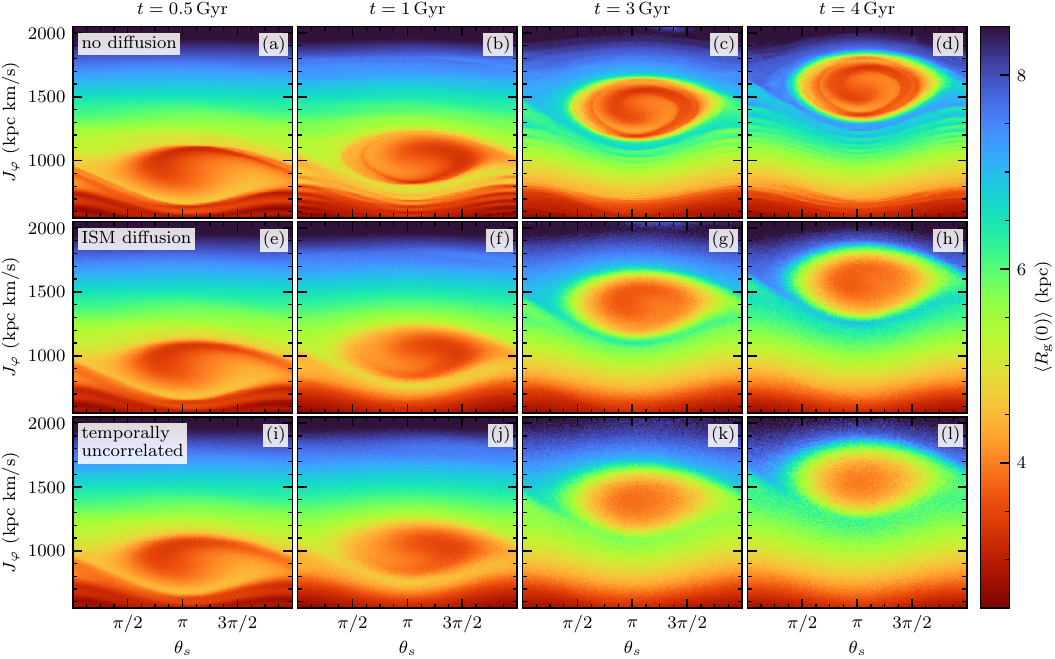}
    \caption{``Tree rings'' at the corotation resonance of a slowing bar \citep{chiba_tree-ring_2021}. Colours show the initial
    guiding radius $R_{\mathrm{g},0}$ of the stars found at each point of the plane
    $(\theta_s,J_\varphi)$ at four times, which increase from left to right. Here,
    $\theta_s=2(\theta_\varphi-\varphi_\mathrm{b})$, $\varphi_\mathrm{b}(t)=\int_0^t
    \Omega_\mathrm{p}(t')\,\md t'$ is the azimuth of the bar, and $\Omega_\mathrm{p}$ is its pattern speed. The top row shows an integration with no diffusion, the middle row one with ISM-driven diffusion, and the bottom row one in which the long-wavelength diffusion coefficients are applied at all $J_R$ and $J_z$, corresponding to impulsive velocity kicks that are uncorrelated in time.}
    \label{fig:bar_example}
\end{figure*}
\begin{figure*}
    \centering
    \includegraphics[width=0.82\linewidth]{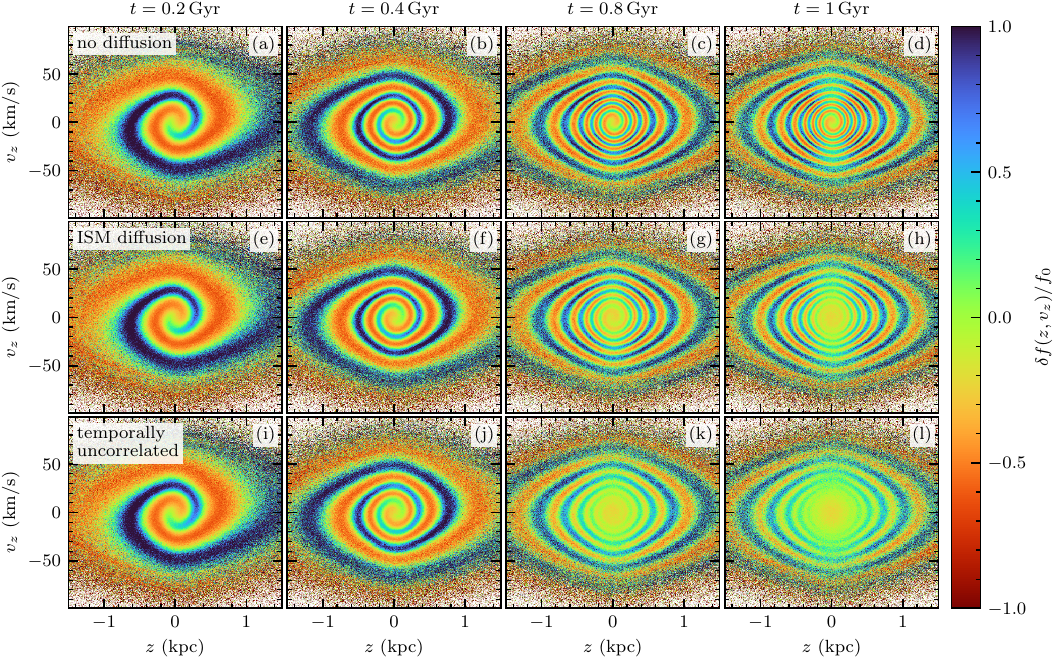}
    \caption{Evolution of a phase spiral in the vertical phase space $(z,v_z)$ of stars near the Sun \citep{2023A&A...673A.115A}. Colours show the fractional perturbation $\delta f/f_0$ to the distribution function, where $\delta f\equiv f-f_0$, at four times, which increase from left to right.
    The spiral is excited by a single kick in $v_z$ at $t=0$, and $f_0$ is the phase-mixed distribution function just before that kick. The rows are as in Figure~\ref{fig:bar_example}.}
    \label{fig:snail}
\end{figure*}

First, consider the `tree rings' at the corotation resonance of the bar, which become visible when one colours cells in the slow angle--angular momentum plane by the mean initial guiding radius, $\langle R_\mathrm{g}(0)\rangle$, of the stars occupying each cell. The initial guiding radius serves as a proxy for metallicity. These structures form because, as the bar slows down, its corotation resonance sweeps outwards through the disc, capturing stars onto librating orbits as it goes. In the absence of diffusion, the order of capture is largely preserved, with stars captured earliest, having the smallest $R_\mathrm{g}(0)$, concentrated near the core of the librating region, and those captured later, with larger $R_\mathrm{g}(0)$, near its edges. ISM-driven diffusion can blur this ordering by mixing stars with different initial guiding radii within the same region of phase space.

To reproduce this feature, and test it in the presence of ISM-driven diffusion, we adopt the axisymmetric \texttt{MilkyWayPotential2022} of \texttt{gala}, combined with the slowing bar of \citet{chiba_resonance_2021}. We initialise a planar ensemble of stellar orbits with a fixed radial action of $J_R=10$ kpc km s$^{-1}$, following the example given in figure 3 of \citet{chiba_tree-ring_2021}. At corotation, $J_R$ corresponds to the fast action, which is approximately conserved during resonant evolution in the absence of diffusion. Fixing its initial value therefore allows us to isolate the resonant dynamics within a single fast-action slice, avoiding the additional broadening that would arise from superposing orbits with different radial actions. We draw the initial guiding radii from an exponential stellar disc with a surface-density scale length of $R_\mathrm{d}=2.5$ kpc, such that $p(\Rg) \md \Rg \propto \Rg \exp(-\Rg/R_\mathrm{d}) \md\Rg$ \citep{juric2008,BH_gerhard2016}, over $2\,\mathrm{kpc}\leq \Rg\leq 10$\,kpc, and sample the initial orbital angles uniformly. The bar initially rotates with a pattern speed of $\Omega_{\rm p,0}=55$ km s$^{-1}$ kpc$^{-1}$ and slows according to $\Omega_{\rm p}(t)=\Omega_{\rm p,0}/(1+\eta\Omega_{\rm p,0}t)$, with $\eta=0.003$. We keep the dimensionless bar strength fixed at $A=0.02$. For the ISM-driven diffusion, we adopt the radially varying gas surface-density profile $\overline{\Sigma}(\Rg)$ of \citet{mcmillan2017}, while keeping the characteristic fluctuation scale $\lambda_*=600$\,pc and correlation time $\tau_*=70$\,Myr fixed throughout the disc and over the course of the integration.

In Figure~\ref{fig:bar_example}, we colour each cell in the plane of angular momentum $J_\varphi$ and slow angle $\theta_s\equiv2(\theta_\varphi-\varphi_\mathrm{b})$ by the mean initial guiding radius, $\langle R_\mathrm{g}(0)\rangle$, of the stars occupying that cell, where $\varphi_\mathrm{b}(t)=\int_0^t\Omega_\mathrm{p}(t'),\md t'$ is the azimuth of the bar and $\Omega_\mathrm{p}$ is its pattern speed. This angle measures the azimuth of a star's guiding centre relative to the bar (the factor of two reflects the twofold symmetry of the bar). It changes slowly for stars near corotation, and stars that are trapped at the resonance oscillate in it, so in the plane $(\theta_s,J_\varphi)$ the trapped region appears as an island. Without diffusion, the trapped region shows sharp nested rings at 3 and 4\,Gyr. With ISM-driven diffusion the trapped region survives, and the stars at its core still come from smaller radii than those at its edge, but the individual rings are blurred. With uncorrelated kicks the blurring is stronger and the edge of the trapped region is less distinct.
The model of \citet{chiba_tree-ring_2021} had no stochastic scattering, and they expected diffusion to weaken the gradient across the trapped region. \citet{chiba_treering_nbody2026} since found that the rings remain well preserved in $N$-body simulations with spiral arms and satellite encounters, which have no ISM. \citet{hamilton_galactic_2023} did include diffusion at a bar resonance, with a constant diffusion coefficient, but for a bar of constant pattern speed, which makes no tree rings. Our example has both a slowing bar and diffusion coefficients that depend on the actions as those of the ISM do, and its outcome is the weakening that \citet{chiba_tree-ring_2021} expected.

Secondly, consider the vertical phase space snail. This forms when a group of stars is displaced from vertical equilibrium. The vertical frequency of a star depends on the amplitude of its oscillation, so the displaced stars wind up into a spiral in the $(z,v_z)$ plane \citep{binney_origin_2018}. To generate such a spiral, we use \texttt{gala} to initialise a stellar population in the Solar neighbourhood in the \texttt{MilkyWayPotential2022} potential. We initialise the planar motion at $R=8.2$ kpc with $v_R=20$ km/s and $v_\varphi=v_\mathrm{circ}(R)$, corresponding to a radial action of approximately $J_R\simeq6$ kpc km s$^{-1}$. Following \citet{gilman_snail2025}, we adopt the vertical distribution function
\begin{equation}
    f(J_z)=n_0\left(1+\frac{J_z\nu}{\alpha\sigma_z^2}\right)^{-\alpha},
\end{equation}
where $n_0$ is a normalisation constant. We adopt their values of $\sigma_z=15.20$ km/s and $\alpha=2.34$, obtained by fitting the vertical velocity dispersion profile measured from Gaia DR3. To approximate an equilibrium population, we integrate the orbits for 2\,Gyr and sample their angles at random times within this interval. We then apply a single vertical velocity kick of $\Delta v_z=10$ km/s to all particles, defining this instant as $t=0$, and evolve the resulting population under different diffusion conditions.

In Figure~\ref{fig:snail} we plot the resulting perturbation to the distribution function at four times. Without diffusion the spiral winds ever more tightly and keeps its full contrast for 1\,Gyr. With ISM-driven diffusion the spiral is still sharp at 0.4\,Gyr, but by 1\,Gyr its inner windings have been erased and only the outer ones remain. With uncorrelated kicks the inner windings are already lost at 0.8\,Gyr, and at 1\,Gyr almost nothing of the spiral is left.

Our kicks have the same form as those with which \citet{tremaine_snail2023} modelled the erasure of the snail by small-scale structure, namely independent random steps of equal variance in the two coordinates of the plane of the vertical oscillation. \citet{chiba_snail2025} used the same kicks, and \citet{gilman_snail2025} used a smoothing of phase space that is calibrated to them. The difference lies in the strength of the kicks. In those papers the strength is a single constant, chosen so that the kicks account for the present vertical velocity dispersion of the disc. This makes $D_{J_zJ_z}$ proportional to $J_z$ at every $J_z$, as it is for forcing that is uncorrelated in time in the vertical long-wavelength regime. The strength of our kicks is set by the diffusion coefficients measured in a simulation of the ISM, which decrease at large $J_R$ and $J_z$. \citet{tremaine_snail2023} found that their kicks erase a snail in about 0.6\,Gyr. In Figure~\ref{fig:snail} the kicks of the bottom row, which like theirs correspond to forcing that is uncorrelated in time, have erased most of the spiral by 1\,Gyr, whereas the realistic ISM-driven kicks of the middle row still leave its outer windings in place.

Two key caveats apply to our idealized model of the snail. Firstly, its stars reach heights $|z|\sim1$\,kpc, far above the thin layer of gas that drives the diffusion, so they are kicked mainly as they cross the midplane. The statistics of their kicks then depend on the angle $\theta_z$, contrary to the assumption of \S\ref{sec:framework}, and our kicks may diffuse $\theta_z$ too fast. Secondly, our recipe treats the vertical oscillation as harmonic when converting from position-velocity to angle-action coordinates (Equation~\ref{eq:Jz}), which is a poor approximation at such heights.

In both examples, then, ISM-driven diffusion erases fine-grained structure within a few Gyr or less, although more slowly than uncorrelated kicks do.

%%%%%%%%%%%%%%%%%%%%%%%%%%%%%%%%%%%%%%%%%%%%%%%%%%%%%%%%%%
%%%%%%%%%%%%%%%%%%%%%%%%%%%%%%%%%%%%%%%%%%%%%%%%%%%%%%%%%%
\subsection{Limitations and extensions}
\label{sec:beyond}
%%%%%%%%%%%%%%%%%%%%%%%%%%%%%%%%%%%%%%%%%%%%%%%%%%%%%%%%%%
%%%%%%%%%%%%%%%%%%%%%%%%%%%%%%%%%%%%%%%%%%%%%%%%%%%%%%%%%%

Our recipe relies on the epicyclic approximation, which is accurate to first order in $a_R/\Rg$ (\S\ref{sec:framework}). For the stars of \S\ref{sec:ism} this ratio stays small: after 10\,Gyr of ISM-driven heating their rms epicyclic amplitude is $\simeq0.65$\,kpc, or $a_R/\Rg\simeq0.08$, and the test in a global potential showed no sign of trouble (\S\ref{sec:box}). For stars on more eccentric orbits, such as those of the thick disc, the epicyclic mapping between $(\bm{r},\bm{v})$ and $(\btheta,\bJ)$ would have to be replaced by a more accurate one, such as the St\"ackel approximation or the other methods for computing actions that are implemented in \texttt{gala}, \texttt{galpy} and \texttt{AGAMA} \citep{pricewhelan2017,bovy2015,vasiliev2019}. The recipe itself would not change, but each kick would then require an evaluation of the actions and angles of the star, which is more expensive than the epicyclic formulae. The recipe would still be fast, and much faster than running a hydrodynamical simulation, for two reasons. Firstly, the kicks can be infrequent: an interval of 50\,Myr between kicks changes the rms changes in the actions by less than 5 per cent (Appendix~\ref{app:convergence}), which amounts to 200 kicks per star over 10\,Gyr. Secondly, the orbit-averaged coefficients depend only on the actions, so they could be tabulated once for a given potential.

Our kicks reproduce the diffusion of the actions by construction. They reproduce the diffusion of the angles exactly only for fluctuations whose force is uniform across the epicycle of a star or uncorrelated in time. The ISM satisfies neither of these criteria precisely, and yet the angle space diffusion is still recovered very well (Figure \ref{fig:box_angles}). On the other hand we have not tested the recipe on stars that travel far above the gas layer, such as those of the phase spiral in \S\ref{sec:illustrations}. For such stars the statistics of the kicks depend on $\theta_z$ and the vertical oscillation is not harmonic, so we leave them to future work (see also Modak et al., in preparation). Another downside to our recipe is that it relies upon a  set of diffusion coefficients being measured and calibrated prior to the simulation, in contrast to modern global approaches that simulate hydrodynamics and $N$-body dynamics self-consistently \citep{teppergarcia2024,teppergarcia2025}. Relatedly, the TIGRESS-NCR simulations with which we calibrated our ISM coefficients do not include any non-axisymmetric gravitational forces other than that due to the ISM (e.g., due to stellar spirals or a bar). It is possible that the ISM in galaxies with global nonaxisymmetries may behave differently (e.g., \citealt{kim2021star}).

Nothing in \S\ref{sec:framework} is specific to the ISM. Our implementation in \texttt{gala} accepts any action-space diffusion coefficients, so the same kicks can include other sources of stochastic forcing, such as transient spiral arms, dark matter substructure or the disc's own self-gravitating fluctuations, once their diffusion coefficients have been measured or calculated. For such forcing the diffusion tensor will not in general be diagonal, as we have taken it to be. The kicks in the three actions are then correlated; they can be drawn with the rule of Equation~\eqref{eq:EM_general}, although the coefficients of Appendix~\ref{app:prescriptions} would gain extra terms that we have not written out explicitly. Kicks of this kind will be useful for interpreting \textit{Gaia} DR4, which will map the phase space of the Galactic disc for more stars and with better precision than DR3. Models of those data will have to include coherent perturbations, such as the bar and spiral arms, alongside stochastic ones such as the ISM. The kicks of this paper offer a cheap way to include the latter in standard orbit integrations, and hence to separate the effects of the two kinds of perturbation on the structures that DR4 will reveal.

%%%%%%%%%%%%%%%%%%%%%%%%%%%%%%%%%%%%%%%%%%%%%%%%%%%%%%%%%%
%%%%%%%%%%%%%%%%%%%%%%%%%%%%%%%%%%%%%%%%%%%%%%%%%%%%%%%%%%
\section{Summary}
\label{sec:summary}
%%%%%%%%%%%%%%%%%%%%%%%%%%%%%%%%%%%%%%%%%%%%%%%%%%%%%%%%%%
%%%%%%%%%%%%%%%%%%%%%%%%%%%%%%%%%%%%%%%%%%%%%%%%%%%%%%%%%%

In this paper we developed a recipe for including ISM-driven orbital diffusion in standard orbit integrations of disc stars. Our work can be summarized as follows:
\begin{itemize}
    \item We showed how to turn a prescribed diffusion in action space into stochastic kicks in position and velocity. The recipe takes as input the three diffusion coefficients $D_{J_\varphi J_\varphi}$, $D_{J_RJ_R}$ and $D_{J_zJ_z}$, and we implemented it in \texttt{gala}.
    \item We tested the recipe in a local shearing box and in a global Milky Way potential. We found that in both cases the kicked orbits reproduce the heating and migration of the orbits that \citet{modak_interstellar_2026} integrated through the multi-physics TIGRESS-NCR simulations of the ISM, to within a few tens of per cent, and often much better, over the full 10\,Gyr of the simulation.
    \item We showed that kicks whose statistics do not depend on the angles of a star make the angles diffuse as well, at a rate fixed by the diffusion of the actions. Our recipe also reproduces the measured angle-space diffusion (but see Appendix~\ref{app:convergence}), despite not requiring explicit angle-space diffusion coefficients as input.
    \item We showed that kicks in velocity alone, with statistics that do not depend on the angles, cannot reproduce ISM-driven diffusion, because they heat stellar orbits too fast once the radial velocity dispersion exceeds about 17\,km\,s$^{-1}$.
    \item We applied the recipe to simple models of the tree rings at the bar's corotation resonance and of the \textit{Gaia} phase spiral. We found that ISM-driven diffusion blurs these structures within a few Gyr, although more slowly than kicks that are uncorrelated in time.
\end{itemize}
Although we focused on the ISM, the recipe applies to any stochastic forcing of stars on nearly circular orbits whose diffusion coefficients in action space are known. The recipe therefore offers a cheap way to include stochastic forcing alongside coherent perturbations, such as the bar and spiral arms, in models of the fine-grained phase-space structure
that \textit{Gaia} DR4 will reveal.

\section*{Acknowledgements}

We thank Eve Ostriker and Scott Tremaine for helpful conversations. We used the large language model Claude (Anthropic) to help with code development and text editing, and we take full responsibility for the content of this paper. This work made use of computational systems managed and supported by Princeton University’s Research Computing. The Flatiron Institute is a division of the Simons Foundation.

\section*{Data Availability}

The stochastic kick recipe described in this paper is implemented in \texttt{gala} \citep[v1.13;][]{pricewhelan2017}. The code and data necessary to reproduce all figures in this paper are publicly available on Zenodo at \href{https://doi.org/10.5281/zenodo.24662274}{10.5281/zenodo.24662274}.

\bibliographystyle{mnras}
\bibliography{main}

@ARTICLE{Kim2023,
       author = {{Kim}, Chang-Goo and {Kim}, Jeong-Gyu and {Gong}, Munan and {Ostriker}, Eve C.},
        title = "{Introducing TIGRESS-NCR. I. Coregulation of the Multiphase Interstellar Medium and Star Formation Rates}",
      journal = {\apj},
         year = 2023,
        month = mar,
       volume = {946},
       number = {1},
          eid = {3},
        pages = {3},
          doi = {10.3847/1538-4357/acbd3a},
archivePrefix = {arXiv},
       eprint = {2211.13293},
 primaryClass = {astro-ph.GA},
       adsurl = {https://ui.adsabs.harvard.edu/abs/2023ApJ...946....3K}
}

@ARTICLE{chiba_treering_nbody2026,
  title={Tree-ring structure of Galactic bar resonance in N-body simulations},
  author={Chiba, Rimpei and Fujii, Michiko and Baba, Junichi and Dubinski, John and Sch{\"o}nrich, Ralph},
  journal={\apj},
  volume={1006},
  number={2},
  pages={132},
  year={2026},
  publisher={The American Astronomical Society}
}

@article{chiba2025galactic,
       author = {{Chiba}, Rimpei and {Ding}, Jupiter and {Hamilton}, Chris and {Kunz}, Matthew W. and {Tremaine}, Scott},
        title = "{Galactic echoes}",
      journal = {\mnras},
         year = 2025,
        month = oct,
       volume = {543},
       number = {1},
        pages = {190-201},
          doi = {10.1093/mnras/staf1463},
archivePrefix = {arXiv},
       eprint = {2506.16512},
 primaryClass = {astro-ph.GA}
}

@article{lacey1983kinematical,
  title={Kinematical and chemical evolution of the galactic disc},
  author={Lacey, CG and Fall, S Michael},
  journal={\mnras},
  volume={204},
  number={3},
  pages={791--810},
  year={1983},
  publisher={Oxford University Press, Oxford, UK}
}

@ARTICLE{mcmillan2017,
       author = {{McMillan}, Paul J.},
        title = "{The mass distribution and gravitational potential of the Milky Way}",
      journal = {\mnras},
         year = 2017,
        month = feb,
       volume = {465},
       number = {1},
        pages = {76-94},
          doi = {10.1093/mnras/stw2759},
archivePrefix = {arXiv},
       eprint = {1608.00971},
 primaryClass = {astro-ph.GA},
       adsurl = {https://scixplorer.org/abs/2017MNRAS.465...76M}
}

@article{hamilton_galactic_2023,
    title = {Galactic {Bar} {Resonances} with {Diffusion}: {An} {Analytic} {Model} with {Implications} for {Bar}-{Dark} {Matter} {Halo} {Dynamical} {Friction}},
    volume = {954},
    issn = {0004-637X},
    shorttitle = {Galactic {Bar} {Resonances} with {Diffusion}},
    url = {https://ui.adsabs.harvard.edu/abs/2023ApJ...954...12H},
    doi = {10.3847/1538-4357/acd69b},
    urldate = {2026-03-19},
    journal = {\apj},
    publisher = {IOP},
    author = {Hamilton, Chris and Tolman, Elizabeth A. and Arzamasskiy, Lev and Duarte, Vinícius N.},
    month = sep,
    year = {2023},
    note = {ADS Bibcode: 2023ApJ...954...12H},
    pages = {12},
}

@article{modak_interstellar_2026,
    title = {Interstellar Medium-Driven Orbital Transport -- I. Radial Heating and Migration},
    url = {http://arxiv.org/abs/2605.21579},
    doi = {10.48550/arXiv.2605.21579},
    urldate = {2026-06-29},
    publisher = {arXiv},
    journal = {arXiv e-prints},
    pages = {arXiv:2605.21579},
    eprint = {2605.21579},
    archivePrefix = {arXiv},
    author = {Modak, Shaunak and Hamilton, Chris and Ostriker, Eve C. and Tremaine, Scott},
    month = {May},
    year = {2026},
}

@ARTICLE{Modak2026residuals,
       author = {{Modak}, Shaunak and {Hackshaw}, Zoe and {Hawkins}, Keith},
        title = "{An Analytic Model for Stellar Metallicity Gradient Residuals in Cold, Phase-mixed Galactic Disks}",
      journal = {\apjl},
         year = 2026,
        month = sep,
       volume = {1008},
       number = {2},
          eid = {L52},
        pages = {L52},
          doi = {10.3847/2041-8213/ae99e2},
archivePrefix = {arXiv},
       eprint = {2606.14850},
 primaryClass = {astro-ph.GA},
       adsurl = {https://scixplorer.org/abs/2026ApJ..1008L..52M}
}

@ARTICLE{Zhang2026,
       author = {{Zhang}, HanYuan and {Belokurov}, Vasily and {Sanders}, Jason L. and {Evans}, N. Wyn and {Chemaly}, David and {Kawata}, Daisuke and {Funakoshi}, Natsuki and {Frankel}, Neige and {Kane}, Sarah G. and {Koposov}, Sergey E.},
        title = "{Orbital migration and heating history of the Galactic disc: a transition between the bimodal discs}",
      journal = {\mnras},
         year = 2026,
        month = sep,
       volume = {551},
       number = {1},
          doi = {10.1093/mnras/stag1406},
archivePrefix = {arXiv},
       eprint = {2512.09987},
 primaryClass = {astro-ph.GA},
       adsurl = {https://scixplorer.org/abs/2026MNRAS.551g1406Z}
}

@inproceedings{kim2021star,
  title={Star Formation and Arm Spurs/Feathers in Local TIGRESS Simulations of Spiral Galaxies},
  author={Kim, W and Kim, C and Ostriker, E},
  booktitle={American Astronomical Society Meeting Abstracts},
  volume={237},
  pages={148--08},
  year={2021}
}

@article{quinlan1992reliability,
  title={On the reliability of gravitational N-body integrations},
  author={Quinlan, Gerald D and Tremaine, Scott},
  journal={Monthly Notices of the Royal Astronomical Society},
  volume={259},
  number={3},
  pages={505--518},
  year={1992},
  publisher={Oxford University Press, Oxford, UK}
}

@article{sellwood2014,
  title = {Secular evolution in disk galaxies},
  author = {Sellwood, J. A.},
  journal = {Rev. Mod. Phys.},
  volume = {86},
  issue = {1},
  pages = {1--46},
  numpages = {0},
  year = {2014},
  month = {Jan},
  publisher = {American Physical Society},
  doi = {10.1103/RevModPhys.86.1},
  url = {https://link.aps.org/doi/10.1103/RevModPhys.86.1}
}

@ARTICLE{BH_gerhard2016,
       author = {{Bland-Hawthorn}, Joss and {Gerhard}, Ortwin},
        title = "{The Galaxy in Context: Structural, Kinematic, and Integrated Properties}",
      journal = {\araa},
         year = 2016,
        month = sep,
       volume = {54},
        pages = {529-596},
          doi = {10.1146/annurev-astro-081915-023441},
archivePrefix = {arXiv},
       eprint = {1602.07702},
 primaryClass = {astro-ph.GA},
       adsurl = {https://scixplorer.org/abs/2016ARA&A..54..529B}
}

@ARTICLE{juric2008,
       author = {{Juri{\'c}}, Mario and {Ivezi{\'c}}, {\v{Z}}eljko and {Brooks}, Alyson and {Lupton}, Robert H. and {Schlegel}, David and {Finkbeiner}, Douglas and {Padmanabhan}, Nikhil and {Bond}, Nicholas and {Sesar}, Branimir and {Rockosi}, Constance M. and et al.},
        title = "{The Milky Way Tomography with SDSS. I. Stellar Number Density Distribution}",
      journal = {\apj},
         year = 2008,
        month = feb,
       volume = {673},
       number = {2},
        pages = {864-914},
          doi = {10.1086/523619},
archivePrefix = {arXiv},
       eprint = {astro-ph/0510520},
 primaryClass = {astro-ph},
       adsurl = {https://scixplorer.org/abs/2008ApJ...673..864J}
}

@ARTICLE{spitzerschwarzchild_51,
       author = {{Spitzer}, Jr., Lyman and {Schwarzschild}, Martin},
        title = "{The Possible Influence of Interstellar Clouds on Stellar Velocities.}",
      journal = {\apj},
         year = 1951,
        month = nov,
       volume = {114},
        pages = {385},
          doi = {10.1086/145478},
       adsurl = {https://ui.adsabs.harvard.edu/abs/1951ApJ...114..385S}
}

@ARTICLE{spitzerschwarzchild_53,
       author = {{Spitzer}, Jr., Lyman and {Schwarzschild}, Martin},
        title = "{The Possible Influence of Interstellar Clouds on Stellar Velocities. II.}",
      journal = {\apj},
         year = 1953,
        month = jul,
       volume = {118},
        pages = {106},
          doi = {10.1086/145730},
       adsurl = {https://ui.adsabs.harvard.edu/abs/1953ApJ...118..106S}
}

@ARTICLE{Modak2026characterizing,
       author = {{Modak}, Shaunak and {Ostriker}, Eve C. and {Hamilton}, Chris and {Tremaine}, Scott},
        title = "{Characterizing Density and Gravitational Potential Fluctuations of the Interstellar Medium}",
      journal = {\apj},
         year = 2026,
        month = mar,
       volume = {999},
       number = {1},
          eid = {136},
        pages = {136},
          doi = {10.3847/1538-4357/ae3bd4},
archivePrefix = {arXiv},
       eprint = {2506.17387},
 primaryClass = {astro-ph.GA},
       adsurl = {https://ui.adsabs.harvard.edu/abs/2026ApJ...999..136M}
}

@ARTICLE{binney_lacey1988,
       author = {{Binney}, James and {Lacey}, Cedric},
        title = "{The diffusion of stars through phase space}",
      journal = {\mnras},
         year = 1988,
        month = feb,
       volume = {230},
        pages = {597-627},
          doi = {10.1093/mnras/230.4.597},
       adsurl = {https://ui.adsabs.harvard.edu/abs/1988MNRAS.230..597B}
}

@article{chiba_tree-ring_2021,
    title = {Tree-ring structure of {Galactic} bar resonance},
    volume = {505},
    copyright = {https://academic.oup.com/journals/pages/open\_access/funder\_policies/chorus/standard\_publication\_model},
    issn = {0035-8711, 1365-2966},
    url = {https://academic.oup.com/mnras/article/505/2/2412/6237521},
    doi = {10.1093/mnras/stab1094},
    language = {en},
    number = {2},
    urldate = {2026-10-08},
    journal = {\mnras},
    author = {Chiba, Rimpei and Schönrich, Ralph},
    month = jun,
    year = {2021},
    pages = {2412--2426},
}

@article{binney_origin_2018,
    title = {The origin of the {Gaia} phase-plane spiral},
    volume = {481},
    copyright = {https://academic.oup.com/journals/pages/open\_access/funder\_policies/chorus/standard\_publication\_model},
    issn = {0035-8711, 1365-2966},
    url = {https://academic.oup.com/mnras/article/481/2/1501/5088367},
    doi = {10.1093/mnras/sty2378},
    language = {en},
    number = {2},
    urldate = {2026-09-18},
    journal = {\mnras},
    author = {Binney, James and Schönrich, Ralph},
    month = dec,
    year = {2018},
    pages = {1501--1506},
}

@ARTICLE{2023A&A...673A.115A,
       author = {{Antoja}, T. and {Ramos}, P. and {Garc{\'\i}a-Conde}, B. and {Bernet}, M. and {Laporte}, C.~F.~P. and {Katz}, D.},
        title = "{The phase spiral in Gaia DR3}",
      journal = {\aap},
         year = 2023,
        month = may,
       volume = {673},
          eid = {A115},
        pages = {A115},
          doi = {10.1051/0004-6361/202245518},
archivePrefix = {arXiv},
       eprint = {2212.11987},
 primaryClass = {astro-ph.GA},
       adsurl = {https://ui.adsabs.harvard.edu/abs/2023A&A...673A.115A}
}

@BOOK{gardiner_stochastic_2009,
       author = {{Gardiner}, Crispin W.},
        title = "{Stochastic Methods: A Handbook for the Natural and Social Sciences}",
      edition = {4th},
    publisher = {Springer},
      address = {Berlin},
         year = 2009,
         isbn = {978-3-540-70712-7}
}

@BOOK{binney_tremaine2008,
       author = {{Binney}, James and {Tremaine}, Scott},
        title = "{Galactic Dynamics: Second Edition}",
    publisher = {Princeton University Press},
      address = {Princeton, NJ},
         year = 2008,
       adsurl = {https://ui.adsabs.harvard.edu/abs/2008gady.book.....B}
}

@ARTICLE{mackereth2019,
       author = {{Mackereth}, J. Ted and {Bovy}, Jo and {Leung}, Henry W. and {Schiavon}, Ricardo P. and {Trick}, Wilma H. and {Chaplin}, William J. and {Cunha}, Katia and {Feuillet}, Diane K. and {Majewski}, Steven R. and {Martig}, Marie and {Miglio}, Andrea and {Nidever}, David and {Pinsonneault}, Marc H. and {Aguirre}, Victor Silva and {Sobeck}, Jennifer and {Tayar}, Jamie and {Zasowski}, Gail},
        title = "{Dynamical heating across the Milky Way disc using APOGEE and Gaia}",
      journal = {\mnras},
         year = 2019,
        month = oct,
       volume = {489},
       number = {1},
        pages = {176-195},
          doi = {10.1093/mnras/stz1521},
archivePrefix = {arXiv},
       eprint = {1901.04502},
 primaryClass = {astro-ph.GA},
       adsurl = {https://ui.adsabs.harvard.edu/abs/2019MNRAS.489..176M}
}

@ARTICLE{sellwood_binney2002,
       author = {{Sellwood}, J.~A. and {Binney}, J.~J.},
        title = "{Radial mixing in galactic discs}",
      journal = {\mnras},
         year = 2002,
        month = nov,
       volume = {336},
       number = {3},
        pages = {785-796},
          doi = {10.1046/j.1365-8711.2002.05806.x},
archivePrefix = {arXiv},
       eprint = {astro-ph/0203510},
 primaryClass = {astro-ph},
       adsurl = {https://ui.adsabs.harvard.edu/abs/2002MNRAS.336..785S}
}

@ARTICLE{frankel2020,
       author = {{Frankel}, Neige and {Sanders}, Jason and {Ting}, Yuan-Sen and {Rix}, Hans-Walter},
        title = "{Keeping It Cool: Much Orbit Migration, yet Little Heating, in the Galactic Disk}",
      journal = {\apj},
         year = 2020,
        month = jun,
       volume = {896},
       number = {1},
          eid = {15},
        pages = {15},
          doi = {10.3847/1538-4357/ab910c},
archivePrefix = {arXiv},
       eprint = {2002.04622},
 primaryClass = {astro-ph.GA},
       adsurl = {https://ui.adsabs.harvard.edu/abs/2020ApJ...896...15F}
}

@ARTICLE{lacey1984,
       author = {{Lacey}, C.~G.},
        title = "{The influence of massive gas clouds on stellar velocity dispersions in galactic discs}",
      journal = {\mnras},
         year = 1984,
        month = jun,
       volume = {208},
        pages = {687-707},
          doi = {10.1093/mnras/208.4.687},
       adsurl = {https://ui.adsabs.harvard.edu/abs/1984MNRAS.208..687L}
}

@ARTICLE{hanninen_flynn2002,
       author = {{H{\"a}nninen}, Jyrki and {Flynn}, Chris},
        title = "{Simulations of the heating of the Galactic stellar disc}",
      journal = {\mnras},
         year = 2002,
        month = dec,
       volume = {337},
       number = {2},
        pages = {731-742},
          doi = {10.1046/j.1365-8711.2002.05956.x},
archivePrefix = {arXiv},
       eprint = {astro-ph/0208426},
 primaryClass = {astro-ph},
       adsurl = {https://ui.adsabs.harvard.edu/abs/2002MNRAS.337..731H}
}

@ARTICLE{aumer2016,
       author = {{Aumer}, Michael and {Binney}, James and {Sch{\"o}nrich}, Ralph},
        title = "{Age-velocity dispersion relations and heating histories in disc galaxies}",
      journal = {\mnras},
         year = 2016,
        month = oct,
       volume = {462},
       number = {2},
        pages = {1697-1713},
          doi = {10.1093/mnras/stw1639},
archivePrefix = {arXiv},
       eprint = {1607.01972},
 primaryClass = {astro-ph.GA},
       adsurl = {https://ui.adsabs.harvard.edu/abs/2016MNRAS.462.1697A}
}

@ARTICLE{hamilton_galactokinetics_2026,
       author = {{Hamilton}, Chris and {Modak}, Shaunak and {Tremaine}, Scott},
        title = "{Galactokinetics}",
      journal = {\apj},
         year = 2026,
        month = jan,
       volume = {997},
       number = {1},
          eid = {28},
        pages = {28},
          doi = {10.3847/1538-4357/ae17bb},
archivePrefix = {arXiv},
       eprint = {2408.03366},
 primaryClass = {astro-ph.GA},
       adsurl = {https://ui.adsabs.harvard.edu/abs/2026ApJ...997...28H}
}

@ARTICLE{hamilton_fouvry2024,
       author = {{Hamilton}, Chris and {Fouvry}, Jean-Baptiste},
        title = "{Kinetic theory of stellar systems: A tutorial}",
      journal = {Phys. Plasmas},
         year = 2024,
        month = dec,
       volume = {31},
       number = {12},
          eid = {120901},
        pages = {120901},
          doi = {10.1063/5.0204214},
archivePrefix = {arXiv},
       eprint = {2402.13322},
 primaryClass = {astro-ph.GA},
       adsurl = {https://ui.adsabs.harvard.edu/abs/2024PhPl...31l0901H}
}

@ARTICLE{teppergarcia2024,
       author = {{Tepper-Garc{\'\i}a}, Thor and {Bland-Hawthorn}, Joss and {Vasiliev}, Eugene and {Agertz}, Oscar and {Teyssier}, Romain and {Federrath}, Christoph},
        title = "{NEXUS: a framework for controlled simulations of idealized galaxies}",
      journal = {\mnras},
         year = 2024,
        month = nov,
       volume = {535},
       number = {1},
        pages = {187-206},
          doi = {10.1093/mnras/stae2372},
archivePrefix = {arXiv},
       eprint = {2406.00342},
 primaryClass = {astro-ph.GA},
       adsurl = {https://ui.adsabs.harvard.edu/abs/2024MNRAS.535..187T}
}

@ARTICLE{teppergarcia2025,
       author = {{Tepper-Garc{\'\i}a}, Thor and {Bland-Hawthorn}, Joss and {Bedding}, Timothy R. and {Federrath}, Christoph and {Agertz}, Oscar},
        title = "{Galactic seismology: can the Gaia 'phase spiral' co-exist with a clumpy, turbulent interstellar medium?}",
      journal = {\mnras},
         year = 2025,
        month = sep,
       volume = {542},
       number = {3},
        pages = {1987-2003},
          doi = {10.1093/mnras/staf1331},
archivePrefix = {arXiv},
       eprint = {2506.15059},
 primaryClass = {astro-ph.GA},
       adsurl = {https://ui.adsabs.harvard.edu/abs/2025MNRAS.542.1987T}
}

@ARTICLE{roskar2008,
       author = {{Ro{\v{s}}kar}, Rok and {Debattista}, Victor P. and {Quinn}, Thomas R. and {Stinson}, Gregory S. and {Wadsley}, James},
        title = "{Riding the Spiral Waves: Implications of Stellar Migration for the Properties of Galactic Disks}",
      journal = {\apjl},
         year = 2008,
        month = sep,
       volume = {684},
       number = {2},
        pages = {L79},
          doi = {10.1086/592231},
archivePrefix = {arXiv},
       eprint = {0808.0206},
 primaryClass = {astro-ph},
       adsurl = {https://ui.adsabs.harvard.edu/abs/2008ApJ...684L..79R}
}

@ARTICLE{antoja2018,
       author = {{Antoja}, T. and {Helmi}, A. and {Romero-G{\'o}mez}, M. and {Katz}, D. and {Babusiaux}, C. and {Drimmel}, R. and {Evans}, D. W. and {Figueras}, F. and {Poggio}, E. and {Reyl{\'e}}, C. and {Robin}, A. C. and {Seabroke}, G. and {Soubiran}, C.},
        title = "{A dynamically young and perturbed Milky Way disk}",
      journal = {\nat},
         year = 2018,
       volume = {561},
       number = {7723},
        pages = {360-362},
          doi = {10.1038/s41586-018-0510-7}
}

@ARTICLE{bovy2015,
       author = {{Bovy}, Jo},
        title = "{galpy: A python Library for Galactic Dynamics}",
      journal = {\apjs},
         year = 2015,
       volume = {216},
       number = {2},
        pages = {29},
          doi = {10.1088/0067-0049/216/2/29}
}

@ARTICLE{pricewhelan2017,
       author = {{Price-Whelan}, Adrian M.},
        title = "{Gala: A Python package for galactic dynamics}",
      journal = {J. Open Source Softw.},
         year = 2017,
       volume = {2},
       number = {18},
        pages = {388},
          doi = {10.21105/joss.00388}
}

@ARTICLE{vasiliev2019,
       author = {{Vasiliev}, Eugene},
        title = "{AGAMA: action-based galaxy modelling architecture}",
      journal = {\mnras},
         year = 2019,
       volume = {482},
       number = {2},
        pages = {1525-1544},
          doi = {10.1093/mnras/sty2672}
}

@ARTICLE{tremaine_snail2023,
       author = {{Tremaine}, Scott and {Frankel}, Neige and {Bovy}, Jo},
        title = "{The origin and fate of the Gaia phase-space snail}",
      journal = {\mnras},
         year = 2023,
       volume = {521},
       number = {1},
        pages = {114-123},
          doi = {10.1093/mnras/stad577}
}

@ARTICLE{friske_schonrich2019,
       author = {{Friske}, Jennifer K. S. and {Sch{\"o}nrich}, Ralph},
        title = "{More than just a wrinkle: a wave-like pattern in U$_g$ versus L$_z$ from Gaia data}",
      journal = {\mnras},
         year = 2019,
       volume = {490},
       number = {4},
        pages = {5414-5423},
          doi = {10.1093/mnras/stz2951}
}

@ARTICLE{hunt_vasiliev2025,
       author = {{Hunt}, Jason A. S. and {Vasiliev}, Eugene},
        title = "{Milky Way dynamics in light of Gaia}",
      journal = {New Astron. Rev.},
         year = 2025,
       volume = {100},
        pages = {101721},
          doi = {10.1016/j.newar.2024.101721}
}

@ARTICLE{chiba_resonance_2021,
       author = {{Chiba}, Rimpei and {Friske}, Jennifer K. S. and {Sch{\"o}nrich}, Ralph},
        title = "{Resonance sweeping by a decelerating Galactic bar}",
      journal = {\mnras},
         year = 2021,
       volume = {500},
       number = {4},
        pages = {4710-4729},
          doi = {10.1093/mnras/staa3585}
}

@ARTICLE{chiba_snail2025,
       author = {{Chiba}, Rimpei and {Frankel}, Neige and {Hamilton}, Chris},
        title = "{Origin of the two-armed vertical phase spiral in the inner Galactic disc}",
      journal = {\mnras},
         year = 2025,
       volume = {543},
       number = {3},
        pages = {2159-2179},
          doi = {10.1093/mnras/staf1566}
}

@ARTICLE{gilman_snail2025,
       author = {{Gilman}, Daniel and {Bovy}, Jo and {Frankel}, Neige and {Benson}, Andrew},
        title = "{Dark Galactic Subhalos and the Gaia Snail}",
      journal = {\apj},
         year = 2025,
       volume = {980},
       number = {1},
          eid = {24},
        pages = {24},
          doi = {10.3847/1538-4357/ada963}
}

%%%%%%%%%%%%%%%%% APPENDICES %%%%%%%%%%%%%%%%%%%%%
\appendix

\section{Kicks in position and velocity from diffusion in action space}
\label{app:prescriptions}

In this appendix we derive the stochastic recipe that we introduced in \S\ref{sec:kicks}, 
showing how we arrived at the position-velocity drift and diffusion coefficients and update rules presented there. As in the main text, we make three assumptions:
\begin{enumerate}
    \item the orbits are epicyclic, so that Equations~\eqref{eq:Jphi}--\eqref{eq:EOM} hold to first order in $a_R/\Rg$,
    \item the diffusion tensor is diagonal in action space, so that the three actions are kicked independently of each other, and
    \item the statistics of kicks at different time-steps do not depend on the angles $\btheta$ of the star.
\end{enumerate}

In the epicyclic approximation the motion of a star is the sum of three parts: the circular motion of its guiding centre, which is described by $J_\varphi$ and $\theta_\varphi$, and its radial and vertical oscillations about the guiding centre, which are described by $(J_R,\theta_R)$ and $(J_z,\theta_z)$. We kick these three parts separately. Kicks associated with radial and vertical heating modify the corresponding oscillations without moving the guiding centre (\S\ref{app:epicycle_kicks}), whereas changes in $J_\varphi$ move the guiding centre and carry the oscillations with it (\S\ref{app:guiding_kicks}). Finally, we explain how the kicks yield the position-velocity drift and diffusion coefficients given in \S\ref{sec:kicks} (\S\ref{app:coefficients}), and derive the correlated Euler--Maruyama update rules (\S\ref{app:update}).

Throughout, $A$ and $D$ denote drift and diffusion coefficients, and their subscripts label the coordinates in which the drift or diffusion occurs. As in \S\ref{sec:kicks}, $\Delta t_\mr{kick}$ is the interval between kicks, angle brackets denote an average over many realisations of the kicks for a star with given actions and angles, and an overbar denotes an average over the orbital angles at fixed actions.

\subsection{Kicks in the radial and vertical actions}
\label{app:epicycle_kicks}

Consider first the radial motion. According to Equations~\eqref{eq:EOM}, an unperturbed star moves around a circle of radius $(2\kappa J_R)^{1/2}$ in the plane $(\kappa[R-\Rg],v_R)$, with $\theta_R$ specifying its position on the circle. To first order in the kick amplitude, a step along the radius of this circle changes $J_R$, whereas a tangential step changes $\theta_R$.

Let $x\equiv R-\Rg$. At fixed guiding radius, Equations~\eqref{eq:JR} and \eqref{eq:EOM} give the first-order changes in radial action and angle due to kicks in $x$ and $v_R$,
\begin{equation}
    \label{eq:DJ_Dtheta}
    \Delta J_R=\frac{\kappa^2x\,\Delta x+v_R\,\Delta v_R}{\kappa},\qquad
    \Delta\theta_R=\frac{v_R\,\Delta x-x\,\Delta v_R}{2J_R}.
\end{equation}

Our assumption of angle-independent kick statistics allows us to take the variances of the kicks in $x$ and $v_R$ to be independent of $\btheta$. Thus we consider zero-mean kicks, independent of each other, with
\begin{equation}
    \label{eq:kick_variances}
    \kappa^2\langle\Delta x^2\rangle=S_R\,\Delta t_\mr{kick},\qquad
    \langle\Delta v_R^2\rangle=S_{v_R}\,\Delta t_\mr{kick},
\end{equation}
where $S_R$ and $S_{v_R}$ are non-negative quantities with units of velocity squared per unit time. Their individual values are not yet specified: the assumption of angle-independent statistics does not require the variances of the two kicks to be equal.

Substituting $x=-a_R\cos\theta_R$ and $v_R=\kappa a_R\sin\theta_R$ from Equations~\eqref{eq:EOM} into Equations~\eqref{eq:DJ_Dtheta}, we find
\begin{align}
    \label{eq:unequal_JJ}
    \frac{\langle\Delta J_R^2\rangle}{\Delta t_\mr{kick}}&=\frac{2J_R}{\kappa}
    \l(S_R\cos^2\theta_R+S_{v_R}\sin^2\theta_R\r),\\
    \label{eq:unequal_thetatheta}
    \frac{\langle\Delta\theta_R^2\rangle}{\Delta t_\mr{kick}}&=\frac{1}{2\kappa J_R}
    \l(S_R\sin^2\theta_R+S_{v_R}\cos^2\theta_R\r),\\
    \label{eq:unequal_Jtheta}
    \frac{\langle\Delta J_R\,\Delta\theta_R\rangle}{\Delta t_\mr{kick}}&=\frac{S_{v_R}-S_R}{\kappa}
    \sin\theta_R\cos\theta_R.
\end{align}
%We see that, although the kick variances are independent of orbital angle, the resulting ensemble-averaged action--angle diffusion coefficients generally \textit{do} depend on orbital angle. 
%However, provided that the actions change little over one orbital period and that kicks sample the orbital angles uniformly, the long-term diffusion rates are obtained by averaging over $\theta_R$. 
Angle-averaging using $\overline{\cos^2\theta_R}=\overline{\sin^2\theta_R}=1/2$ and $\overline{\sin\theta_R\cos\theta_R}=0$, we find
\begin{align}
    \label{eq:DJJ_sum}
    D_{J_RJ_R}&\equiv\frac{\overline{\langle\Delta J_R^2\rangle}}{\Delta t_\mr{kick}}
    =\frac{J_R}{\kappa}(S_R+S_{v_R}),\\
    \label{eq:Dthetatheta}
    D_{\theta_R\theta_R}&\equiv\frac{\overline{\langle\Delta\theta_R^2\rangle}}{\Delta t_\mr{kick}}
    =\frac{S_R+S_{v_R}}{4\kappa J_R}
    =\frac{D_{J_RJ_R}}{4J_R^2},
\end{align}
and $D_{J_R \theta_R} = 0$.
%The mixed term also averages to zero, $\overline{\langle\Delta J_R\,\Delta\theta_R\rangle}=0$. A correlation between the kicks in $x$ and $v_R$ would contribute additional angle-dependent terms that likewise vanish upon orbit averaging.
%Since successive stochastic kicks are independent, their variances add over time. Provided that many kicks occur per orbital period and the actions vary slowly, 
These angle-averaged coefficients describe the diffusion over many orbital periods. Crucially, both depend only on the sum $S_R+S_{v_R}$: orbit averaging has removed any information about how the diffusion is distributed between the two pieces $S_R$ and $S_{v_R}$.  We are therefore free to choose any combination of $S_R$ and $S_{v_R}$ that has the correct sum. 
%The measured $D_{J_RJ_R}$ therefore constrains $S_R+S_{v_R}$, but not the individual contributions. 
We adopt the simplest choice $S_R=S_{v_R}$, giving equal variances to the kicks in $\kappa R$ and $v_R$:
\begin{equation}
    \label{eq:isotropic_kicks}
    \kappa^2\langle\Delta R^2\rangle=\langle\Delta v_R^2\rangle
    =\frac{\kappa D_{J_RJ_R}}{2J_R}\,\Delta t_\mr{kick}.
\end{equation}

We now turn to the drift. Equations~\eqref{eq:DJ_Dtheta} hold to first order in the kick amplitude only.
To second order, kicks $\Delta x$ and $\Delta v_R$ change the radial action by
\begin{equation}
    \label{eq:DeltaJR}
    \Delta J_R = \frac{\kappa^2x\,\Delta x+v_R\,\Delta v_R}{\kappa}
    +\frac{\kappa^2\Delta x^2+\Delta v_R^2}{2\kappa}.
\end{equation}
For the kicks of Equation~\eqref{eq:isotropic_kicks}, the first term has zero mean and produces the desired diffusion $D_{J_RJ_R}$, while the second induces a drift $D_{J_RJ_R}/(2J_R)$. To recover the drift $A_{J_R}$ prescribed by Equation~\eqref{eq:drift}, we rescale the epicycle at each kick as
$(x,v_R)\mapsto(1+c_R\,\Delta t_\mr{kick})(x,v_R)$. This changes $J_R$ by $2J_Rc_R\,\Delta t_\mr{kick}$, giving

\begin{equation}
    \label{eq:cR}
    c_R = \frac{A_{J_R}}{2J_R}-\frac{D_{J_RJ_R}}{4J_R^2}.
\end{equation}
The corresponding drift terms are $c_R\,x$ and $c_R\,v_R$, as given in \S\ref{sec:kicks}. Note that $c_R=0$ wherever $D_{J_RJ_R}\propto J_R$, since the stochastic kicks themselves induce the required drift.

By construction, the kicks associated with radial heating leave $J_\varphi=Rv_\varphi$ and $\theta_\varphi=\varphi-\gamma v_R/(\kappa\Rg)$ unchanged to first order in both the epicyclic amplitude and the kick amplitudes. Each kick in $R$ therefore comes with a kick $\Delta v_\varphi=-(v_\varphi/R)\Delta R$, and each kick in $v_R$ by a kick $\Delta\varphi=\gamma\Delta v_R/(\kappa\Rg)$. Leaving $\theta_\varphi$ unchanged is a choice: without the latter kick, $v_R$ kicks would induce direct diffusion in $\theta_\varphi$. However, the choice matters little in practice: in the shearing box of \S\ref{sec:box}, the resulting rms change after 1\,Gyr would be at most 0.06\,rad, compared to 0.6\,rad arising indirectly from changes in $\Omega(\Rg)$.

The same construction applies to vertical motion in the plane $(\nu z,v_z)$. At fixed guiding centre, we may introduce independent zero-mean kicks with angle-independent variances
\begin{equation}
    \label{eq:vertical_kick_variances}
    \nu^2\langle\Delta z^2\rangle=S_z\,\Delta t_\mr{kick},\qquad
    \langle\Delta v_z^2\rangle=S_{v_z}\,\Delta t_\mr{kick}.
\end{equation}
Here, $S_z$ and $S_{v_z}$ play the same roles as $S_R$ and $S_{v_R}$ above. Just as in the radial case, only their sum is constrained:
\begin{equation}
    \label{eq:vertical_angle_averages}
    D_{J_zJ_z}=\frac{J_z}{\nu}(S_z+S_{v_z}),\qquad
    D_{\theta_z\theta_z}=\frac{S_z+S_{v_z}}{4\nu J_z}
    =\frac{D_{J_zJ_z}}{4J_z^2}.
\end{equation}
We choose $S_z=S_{v_z}$, which gives
\begin{equation}
    \label{eq:vertical_isotropic_kicks}
    \nu^2\langle\Delta z^2\rangle=\langle\Delta v_z^2\rangle
    =\frac{\nu D_{J_zJ_z}}{2J_z}\,\Delta t_\mr{kick}.
\end{equation}
%As in the radial case, these equal-variance kicks yield the above action and angle diffusion rates at every $\theta_z$, without an orbit average. 
The stochastic kicks also induce a mean drift $D_{J_zJ_z}/(2J_z)$, which need not be the prescribed $A_{J_z}$. We therefore rescale the vertical oscillation as $(z,v_z)\mapsto(1+c_z\,\Delta t_\mr{kick})(z,v_z)$, where
\begin{equation}
    \label{eq:cz}
    c_z=\frac{A_{J_z}}{2J_z}-\frac{D_{J_zJ_z}}{4J_z^2}.
\end{equation}
The corresponding drift terms are $c_z z$ and $c_z v_z$. Unlike radial heating, vertical heating requires no companion kicks to preserve the guiding centre. Since vertical heating is decoupled from the motion in the plane to first order in the epicyclic amplitude, we could instead choose velocity-only kicks, $S_z=0$, while retaining the same orbit-averaged $D_{J_zJ_z}$; our equal-variance choice keeps the vertical and radial constructions consistent.

\subsection{Kicks in the angular momentum}
\label{app:guiding_kicks}

We implement both the stochastic diffusion and the systematic drift in $J_\varphi$ by changing the angular momentum according to Equation~\eqref{eq:EM_actions}. A change $\Delta J_\varphi$ moves the guiding radius by
\begin{equation}
    \label{eq:DeltaRg}
    \Delta\Rg = \frac{\md\Rg}{\md J_\varphi}\Delta J_\varphi
    +\frac{1}{2}\frac{\md^2\Rg}{\md J_\varphi^2}\Delta J_\varphi^2.
%    \qquad \frac{\md\Rg}{\md J_\varphi}=\frac{\gamma^2\Rg}{2J_\varphi}.
\end{equation}
We move the star with its guiding centre, so that $x$, $v_R$, $z$ and $v_z$ do not change. Then $\Delta R=\Delta\Rg$ and, to first order in $\Delta J_\varphi$, $\Delta v_\varphi=(\md v_c/\md\Rg)\,\Delta\Rg=(1-\gamma^2/2)\,\Delta J_\varphi/\Rg$. The second-order term in Equation~\eqref{eq:DeltaRg} contributes to the drift in $R$. A change in guiding radius without a corresponding displacement of the star would also change $x=R-\Rg$ and hence introduce spurious radial heating. Carrying the star with its guiding centre avoids this coupling between migration and radial heating. In particular, although taking $S_R=0$ would still reproduce the required orbit-averaged $D_{J_RJ_R}$ in the absence of migration, a prescription restricted to velocity kicks could not independently implement guiding-radius shifts in this way (see \S\ref{subsec:velocity_kicks}). We choose to leave $\theta_\varphi$ unchanged by the kicks themselves; changes in $\Rg$ nevertheless affect its subsequent evolution by changing the azimuthal frequency $\Omega(\Rg)$.
%To write it out we define the shear parameters
%\begin{equation}
%    q\equiv-\frac{\md\ln\Omega}{\md\ln R},\quad q_R \equiv -\frac{\md\ln\kappa}{\md\ln R},
%\end{equation}
%which give the identity $\gamma^2 = 2/(2-q)$.
%From the definition of the epicyclic frequency it follows that
%\begin{equation}
%    q_R = q\left(1 + \frac{\gamma^2}{4}\frac{\md\ln q}{\md \ln R}\right),
%\end{equation}
%so $q_R = q$ for any power-law rotation curve. For example, in a pure logarithmic halo potential with a constant circular velocity
%$\Rg\Omega=V_0$, $\gamma=\sqrt{2}$ and $q=q_R=1$.

\subsection{Diffusion and drift coefficients}
\label{app:coefficients}

Combining the results of the previous two subsections   gives the diffusion and drift coefficients for $(\varphi,R,z,v_\varphi,v_R,v_z)$ presented in \S\ref{sec:kicks}. Although the action-space diffusion tensor is diagonal, the position-velocity tensor has nonzero cross terms: $D_{Rv_\varphi}$ reflects the companion kicks needed to keep $J_\varphi$ fixed during radial heating, and $D_{\varphi v_R}$ reflects our choice to preserve $\theta_\varphi$. By contrast, the vertical coordinates are decoupled from the other coordinates to leading epicyclic order. The three terms of $A_R$ arise from the mean shift of the guiding radius, the radial rescaling of Equation~\eqref{eq:cR}, and the second-order contribution in Equation~\eqref{eq:DeltaRg}. The drift $A_{v_\varphi}$ follows by requiring that the mean change in $J_\varphi=Rv_\varphi$ be $A_{J_\varphi}\,\Delta t_\mr{kick}$, including the covariance between $R$ and $v_\varphi$. The corresponding vertical drift is the rescaling given by Equation~\eqref{eq:cz}.

Note that the position-velocity drifts $A$ cannot in general be found by differentiating the position-velocity diffusion coefficients $D$ in the manner of Equation~\eqref{eq:drift}. That relation holds for canonical coordinates such as the angles and actions, whereas $(\varphi,R,z,v_\varphi,v_R,v_z)$ are not canonical.

\begin{figure}
    \centering
    \includegraphics[width=0.88\columnwidth]{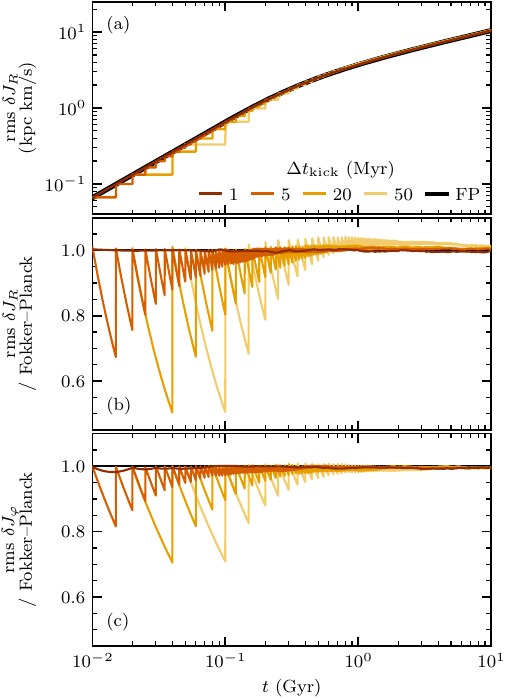}
    \caption{Convergence with the interval between kicks, in the shearing box. Panel (a) shows the rms change in the radial action from a direct integration of the Fokker--Planck equation (black line) and from kicked orbits with intervals $\Delta t_\mr{kick}$ between kicks of 1, 5, 20 and 50\,Myr (vermilion to gold lines, from darker to lighter). Panels (b) and (c) show the rms changes in the radial and azimuthal actions
    of the kicked orbits, divided by those of the direct integration.}
    \label{fig:convergence}
\end{figure}

\begin{figure*}
    \centering
    \includegraphics[width=0.77\textwidth]{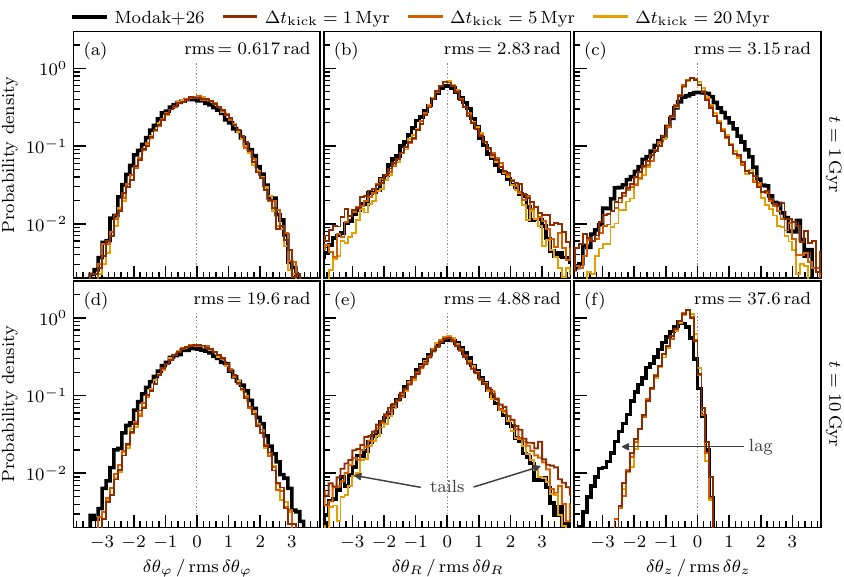}
    \caption{Distributions of the changes in the angles for three intervals between kicks, in the shearing box. From left to right, the columns show the changes in $\theta_\varphi$, $\theta_R$ and $\theta_z$ relative to unperturbed orbits (Equation~\ref{eq:dtheta}), measured from $t_0=100$\,Myr. The top row is at $t=1$\,Gyr and the bottom row at $t=10$\,Gyr. In each panel the changes are divided by their rms value in the simulation of \citet{modak_interstellar_2026} at that time, which is given at the top right. Thick black lines show the orbits integrated through TIGRESS-NCR by \citet{modak_interstellar_2026}. Thin lines show stars kicked in the shearing box with intervals $\Delta t_\mr{kick}$ between kicks of 1, 5 and 20\,Myr (vermilion to gold, from darker to lighter). Arrows mark the tails of $\delta\theta_R$ in panel (e) and the lag of $\theta_z$ in panel (f). In panels (a) and (d) the changes are measured relative to the angular frequency of the centre of the box.}
    \label{fig:angle_hist_dt}
\end{figure*}

\subsection{Update rules}
\label{app:update}

We now have drifts and diffusion coefficients for six coordinates, and we need a rule for kicking a star that reproduces all of them.
%If no two coordinates were coupled, we could update each one as in
%Equation~\eqref{eq:EM_actions}, with one random number per coordinate.
%However, two of our diffusion coefficients couple pairs of coordinates: $D_{Rv_\varphi}$ couples
%$R$ to $v_\varphi$, and $D_{\varphi v_R}$ couples $\varphi$ to $v_R$.
%The kicks within each of these pairs are therefore correlated.
For a set of coordinates $X_a$ with drifts $A_a$ and a diffusion tensor $D_{ab}$ that is not
diagonal, the Euler--Maruyama update rule is \citep[e.g.,][]{gardiner_stochastic_2009}
\begin{equation}
    \label{eq:EM_general}
    X_a \mapsto X_a + A_a\,\Delta t_\mr{kick} + \sum_b L_{ab}\,\eta_b\sqrt{\Delta t_\mr{kick}},
\end{equation}
where the $\eta_b$ are independent random numbers drawn from a normal distribution with zero mean and unit variance, and $L_{ab}$ is any matrix that satisfies $\sum_c L_{ac}L_{bc}=D_{ab}$, such as the Cholesky factor of $D_{ab}$. For a pair of coupled coordinates, which we label 1 and 2, the Cholesky factor has $L_{11}=\sqrt{D_{11}}$, $L_{12}=0$, $L_{21}=D_{12}/\sqrt{D_{11}}$ and $L_{22}=(D_{22}-D_{12}^2/D_{11})^{1/2}$. In other words, the first coordinate of the pair is kicked with one random number, and the second with the same random number and one of its own. We apply this rule to the pair $(R,v_\varphi)$ and to the pair $(v_R,\varphi)$, and we update $z$ and $v_z$, which are coupled to nothing, as in Equation~\eqref{eq:EM_actions}. That would take six random numbers, but only five are needed: for the pair $(v_R,\varphi)$ the coefficients of \S\ref{sec:kicks} satisfy $D_{\varphi\varphi}=D_{\varphi v_R}^2/D_{v_Rv_R}$, so that $L_{22}=0$ and $\varphi$ has no random number of its own. The kick in $\varphi$ is $\gamma/(\kappa\Rg)$ times the kick in $v_R$, as required to keep the guiding centre fixed in azimuth (\S\ref{app:epicycle_kicks}). Writing the five independent random numbers as $\eta_R$, $\eta_{v_\varphi}$, $\eta_{v_R}$, $\eta_z$ and $\eta_{v_z}$ gives the six update rules presented in Equations~\eqref{eqn:realspace_update_1}--\eqref{eqn:realspace_update_6} of \S\ref{sec:kicks}. 
%The two kicks in $(R,v_\varphi)$ share one random number and require a second, whereas the stochastic increments in $(v_R,\varphi)$ are perfectly correlated and share a single random number. The remaining two random numbers independently update $z$ and $v_z$.

\section{Numerical convergence}
\label{app:convergence}

In this appendix we ask two questions of the kicks, both in the shearing box of \S\ref{sec:box}: do the kicked orbits obey the diffusion equation that they are meant to obey? And how long can the interval $\Delta t_\mr{kick}$ between kicks be? To answer the first question, we integrated Equation~\eqref{eq:QL_diffusion} directly on a grid in $J_R$, for stars that start on circular orbits. In panel (a) of Figure~\ref{fig:convergence} we plot rms\,$\delta J_R$ from this direct integration and from kicked orbits with $\Delta t_\mr{kick}=1$, 5, 20 and 50\,Myr. Between kicks the actions do not change, so each of the kicked curves is a staircase. In panels (b) and (c) we divide each staircase by the result of the direct integration, for $J_R$ and for $J_\varphi$.

With $\Delta t_\mr{kick}=1$\,Myr the kicked orbits agree with the direct integration to within 1 per cent at all times. With longer intervals the two still agree immediately after each kick, to within 5 per cent for $\Delta t_\mr{kick}=50$\,Myr and 2 per cent for the shorter intervals. Between kicks each staircase falls behind the direct integration, by a fraction of order $\Delta t_\mr{kick}/t$ at time $t$, and these are the teeth in panels (b) and (c). The interval between kicks therefore only has to be a few times shorter than the earliest time of interest. We found the same in the global potential, with intervals of up to 100\,Myr.

In Figure~\ref{fig:angle_hist_dt} we compare the distributions of the changes in the angles with those of \citet{modak_interstellar_2026} (for $\theta_\varphi$ and $\theta_R$) and Modak et al., in preparation (for $\theta_z$), for $\Delta t_\mr{kick}=1$, 5 and 20\,Myr. For $\theta_\varphi$ the distributions agree at both times, whatever $\Delta t_\mr{kick}$ (panels a and d). For $\theta_R$ the cores agree, but the tails grow as $\Delta t_\mr{kick}$ decreases (arrows in panel e), so that rms\,$\delta\theta_R$ at 10\,Gyr rises from 4.7\,rad for $\Delta t_\mr{kick}=20$\,Myr to 5.7\,rad for 1\,Myr, against 4.9\,rad in the simulation. These tails are made by stars that pass close to $J_R=0$, where even a small kick turns the phase of the epicycle through a large angle. With smaller kicks this continues down to smaller $J_R$, so the tails are heavier. The tails of $\delta\theta_z$ behave in the same way at 1\,Gyr (panel c). At 10\,Gyr, however, $\delta\theta_z$ is dominated by a lag (arrow in panel f), which arises because the vertical frequency of a star falls as its $J_z$ grows, and the exact form of this dependence is sensitive to the vertical structure of the gas layer. The difference in lags between our shearing box model and the direct integrations of Modak et al. (in preparation) arise from the fact that although we have configured our shearing box with a mean gas layer that yields the correct vertical frequency evaluated at $J_z = 0$ (i.e., the correct midplane gas volume density), our simulations do not explicitly match the time-dependent mean vertical structure of the gas in the TIGRESS-NCR R8 simulations (see, e.g., \S4.1 of \citealt{Modak2026characterizing}). We emphasize that the difference in the lag is entirely due to this vertical structure and not our kick model.

%%%%%%%%%%%%%%%%%%%%%%%%%%%%%%%%%%%%%%%%%%%%%%%%%%

% Don't change these lines
\bsp	% typesetting comment
\label{lastpage}
\end{document}